\documentclass[prb,aps,twocolumn,groupedaddress,nofootinbib]{revtex4}
\usepackage[usenames,dvipsnames,svgnames]{xcolor} 
\usepackage{graphicx}
\usepackage[export]{adjustbox}
\usepackage{amsmath,amsfonts,amssymb,latexsym}
\usepackage{hhline}
\usepackage{bm}
\usepackage{verbatim}
\usepackage{enumitem}
\usepackage{mathrsfs}
\usepackage{slashed}
\usepackage{empheq}

\newcommand{\sgn}{{\rm sgn}}

\newcommand{\p}{\partial}

\newcommand{\lan}{\langle}
\newcommand{\ran}{\rangle}

\newcommand{\unit}{\mathbf{1}}

\newcommand{\gp}{{\gamma'}}

\newcommand{\da}{{\dagger}}

\newcommand{\ra}{\rightarrow}
\newcommand{\lra}{\leftrightarrow}

\newcommand{\wt}{\widetilde}

\renewcommand{\(}{\left(}
\renewcommand{\)}{\right)}
\renewcommand{\[}{\left[}
\renewcommand{\]}{\right]}
\newcommand{\mt}{\mapsto}

\newcommand{\dk}{\frac{dk}{2\pi}\,}

\newcommand{\twp}{{2\pi}}
\newcommand{\fpi}{{4\pi}}

\newcommand{\D}{\nabla}

\newcommand\bpm            {\begin{pmatrix}}
\newcommand\epm           {\end{pmatrix}}

\newcommand{\vp}{\varphi}

\newcommand{\cp}{\Phi}
\newcommand{\ct}{\Theta}
\newcommand{\lr}{{L/R}}
\newcommand{\rl}{{R/L}}

\newcommand{\inv}{^{-1}}

\newcommand{\ope}\odot

\renewcommand{\prl}{{\, \parallel \,}}

\newcommand\be            {\begin{equation}}
\newcommand\ee            {\end{equation}}
\newcommand\ba            {\begin{aligned}}
\newcommand\ea            {\end{aligned}}
\newcommand\bea{\begin{equation}\begin{aligned}}
\newcommand\eea{\end{aligned}\end{equation}}

\newcommand{\lap}{{\Lambda_\perp}}
\newcommand{\Lapr}{\Lambda_\parallel}

\usepackage{subcaption}

 \usepackage{hyperref} 
 \hypersetup{final}
 \hypersetup{colorlinks, citecolor=MidnightBlue, linkcolor=MidnightBlue, urlcolor=MidnightBlue}

 \renewcommand{\ss}{\subsection}
 
 \renewcommand{\a}{\alpha}
 \renewcommand{\b}{\beta}
 \renewcommand{\d}{\delta}
 \newcommand{\De}{\Delta}
 \newcommand{\g}{\gamma}
 \newcommand{\G}{\Gamma}
 \newcommand{\s}{\sigma}
 \renewcommand{\S}{\Sigma} 
 
 \newcommand{\ep}{\varepsilon} 
 \renewcommand{\l}{\lambda}
 \renewcommand{\L}{\Lambda}
 \renewcommand{\t}{\theta}
 
 \renewcommand{\o}{\omega}
 \renewcommand{\O}{\Omega}

 \renewcommand{\r}{\rho}
 
 \renewcommand{\c}{\chi}
 \newcommand{\z}{\zeta}

 \newcommand{\bfmu}{{\boldsymbol{\mu}}}

 \newcommand{\bfep}{{\boldsymbol{\epsilon}}}
 \newcommand{\bfga}{{\boldsymbol{\gamma}}}

 \newcommand{\bfA}{\mathbf{A}}

 \newcommand{\bfK}{\mathbf{K}}
 
 \newcommand{\bfM}{\mathbf{M}}

 \newcommand{\bfk}{\mathbf{k}}

 \newcommand{\bfq}{\mathbf{q}}

 \newcommand{\bfx}{\mathbf{x}}
 
 \newcommand{\bfy}{\mathbf{y}}

 \newcommand{\zz}{\mathbb{Z}}

 \newcommand{\mco}{\mathcal{O}}

 \newcommand{\mcl}{\mathcal{L}}
 \newcommand{\mcg}{\mathcal{G}}

 \newcommand{\sfc}{\mathsf{c}}
 \newcommand{\sfd}{\mathsf{d}}

 \newcommand{\sfk}{\mathsf{k}}

 \usepackage[mathscr]{eucal} 

\usepackage{braket}

\usepackage{tikz}

\usepackage{dcolumn}
\newcommand{\dkdo}{\frac{dk \, d\o}{(\twp)^2}}
\newcommand{\dthq}{\frac{d^3 q}{(\twp)^3}}
\newcommand{\bbcs}{\bar g_{BCS}}
\renewcommand{\Lapr}{\L}
\renewcommand{\lap}{\L} 
\renewcommand{\bfep}{{\boldsymbol{\varepsilon}}}

\newcommand{\ob}[1]{\mkern 1.5mu\overline{\mkern-1.5mu#1\mkern-1.5mu}\mkern 1.5mu}

\begin{document}

\title{Bose-Luttinger Liquids}

\author{Ethan Lake}
\author{T. Senthil}
\affiliation{Department of Physics, Massachusetts Institute of Technology, Cambridge, MA, 02139}
\author{Ashvin Vishwanath}
\affiliation{Department of Physics, Harvard University, Cambridge, MA 02138}
\begin{abstract}
	We study systems of bosons whose low-energy excitations are located along a spherical submanifold of momentum space. We argue for the existence of gapless phases which we dub ``Bose-Luttinger liquids'', which in some respects can be regarded as bosonic versions of Fermi liquids, while in other respects exhibit striking differences. These phases have bosonic analogues of Fermi surfaces, and like Fermi liquids they possess a large number of emergent conservation laws. Unlike Fermi liquids however these phases lack quasiparticles, possess different RG flows, and have correlation functions controlled by a continuously varying exponent $\eta$, which characterizes the anomalous dimension of the bosonic field. We show that when $\eta>1$, these phases are stable with respect to all symmetric perturbations. These theories may be of relevance to several physical situations, including frustrated quantum magnets, rotons in superfluid He, and superconductors with finite-momentum pairing. As a concrete application, we show that coupling a Bose-Luttinger liquid to a conventional Fermi liquid produces a resistivity scaling with temperature as $T^\eta$. We argue that this may provide an explanation for the non-Fermi liquid resistivity observed in the paramagnetic phase of MnSi. 
	
\end{abstract}

\maketitle

\tableofcontents

\section{Introduction and summary}

The difficulty of understanding a given phase of matter roughly scales with the number of low-energy degrees of freedom it possesses. Phases with finitely many low-energy degrees of freedom are relatively easy to understand, and can be classified using the framework of topological quantum field theory. More difficult are theories where the gap goes to zero at isolated points in momentum space. The low energy physics of these theories are described by gapless quantum field theories. In many cases these field theories are conformal, and can be understood using powerful techniques from conformal field theory. More difficult still are a third class of theories possessing a larger amount of gapless degrees of freedom, with gapless modes located along a nontrivial submanifold of momentum space. The canonical examples of such theories are Fermi liquids and non-Fermi liquids. 

This third class of ``very gapless'' phases of matter is of fundamental importance to condensed matter physics, but it is at present unclear whether or not phases in this class can be understood within any particular unifying framework. It is therefore valuable to construct examples of such theories beyond the purview of (non-)Fermi liquids, in order to understand what general features such phases of matter are expected to possess.

In this paper, we will study phases of bosons which fall into this third class of matter. The systems we will study have dispersion relations like
\be \label{bll_disp} \ep(\bfk ) \sim \sqrt{r + v^2(k^2-k_B^2)^2},\ee 
so that $\ep(\bfk)$ is degenerate along a sphere of radius $k_B$ in momentum space, which we refer to as a ``Bose surface''.

We will be interested in scenarios in which amplitude ordering occurs across the entire Bose surface. In these scenarios, the phase degrees of freedom at each point on the Bose surface fluctuate in a quasi-one-dimensional manner, preventing the establishment of long-range phase ordering.
In the same way that Fermi liquids can be thought of as a collection of 1+1D Dirac fermions, with one Dirac fermion for each point of the Fermi surface, we will see that these phases can be regarded as collections of 1+1D Luttinger liquids, with one Luttinger liquid located at each point on the Bose surface. As such, we dub these phases ``Bose-Luttinger liquids''.\footnote{Note that such phases are conceptually distinct from ``Bose metals'', viz. systems of bosons (usually Cooper pairs) which at $T=0$ have metallic transport\cite{das1999existence,phillips2003elusive,dalidovich2001interaction,dalidovich2002phase}. We are instead interested in theories that possess a large number of gapless excitations (regardless of whether or not they are metals).} 

Our motivation for studying these types of systems is two-fold.
 First, whether or not such ``very gapless'' phases can arise in purely bosonic systems (without fine-tuning) is an interesting question in its own right, since one cannot rely on degeneracy pressure to stabilize the Bose surface. In fact a similar question has already arisen in the literature, where it appeared in the context of various two-dimensional ring-exchange models.\cite{paramekanti2002ring,ma2018higher,seiberg2020exoticI,seiberg2020exoticII,tay2011possible,xu2007bond,you2020emergent,xu2005reduction} These models have an anisotropic dispersion which vanishes along the coordinate axes in momentum space, and are described in the IR by field theories exhibiting quasi 1+1D behavior. However, the stability of these models in the thermodynamic limit is a rather delicate issue, and may require the presence of a UV symmetry group with an infinite number of conserved charges. By contrast, the phases we will study in this paper are closer in spirit to Fermi liquids --- they are rotation-invariant, and are stable in the presence of a small UV symmetry group consisting only of translation and $U(1)$ charge conservation.
 
 Our second motivation for studying these types of theories can be traced back to an old idea of Anderson,\cite{anderson1990luttinger} who proposed that Fermi liquids in 2+1D are generically unstable, and instead flow in the IR to Luttinger liquid like fixed points that lack well-defined quasiparticles. 
 
 This proposal unfortunately turned out to be incorrect, with the geometry of the Fermi surface protecting the quasiparticle pole against interactions, as long as the interactions are sufficiently non-singular. 
 While interactions are not able to easily create a phase with Luttinger liquid type exponents, this obstacle can be overcome by working instead with systems of bosons, where the Luttinger liquid behavior can be built in at a more fundamental level. 
 We will see how this line of reasoning can be used to construct fixed points that share some similarities to those envisioned by Anderson. However as we will see, there are also significant differences in  the precise structure of the low energy theory, and the underlying degrees of freedom are bosonic, rather than fermionic.  
 
 The Bose-Luttinger liquids studied in this paper are phenomenologically somewhat similar to Fermi liquids, although there are many important differences. Like Fermi liquids these phases are metals, have a $T$-linear specific heat, possess correlation functions exhibiting oscillations at integer multiples of a ``Bose momentum'' $k_B$, and have a set of Landau parameters which modify some aspects of their phenomenology. 
 Unlike Fermi liquids however these phases lack quasiparticles, have correlation functions with continuously tunable exponents, and will be seen to possess rather different RG flows.

The structure of this paper is as follows. In section \ref{sec:oned}, we warm up by considering a simple one-dimensional example of a Bose-Luttinger liquid, which like a one-dimensional Fermi liquid involves a dispersion which is gapless at two ``Bose points'' in momentum space. In the IR this theory can be understood as a multi-component Luttinger liquid enriched with a particular symmetry action.

We then move on to explore a generalization of this example to 2+1D, which is the main focus of this paper. The UV model is introduced in section \ref{sec:twod}, and consists of translation-invariant conserved bosons with a dispersion possessing degenerate minima along a circle in momentum space. In section \ref{sec:2dirtheory} we write down a Lagrangian describing the low-energy physics of the Bose-Luttinger liquid fixed point, and discuss the emergent symmetries and operator content of the IR theory. In these two sections, we assume the presence of a microscopic particle-hole symmetry which fixes the system to be at zero average density. This is done only for simplicity, and in section \ref{sec:finite_density} we explain the generalization to the finite density case. 

In section \ref{sec:2drg} we set up an RG analysis to determine the stability of the Bose-Luttinger fixed point. We find a regime of parameter space where the fixed point is stable against all symmetric perturbations, and another regime where it possesses an instability with respect to a BCS-type pairing interaction. 
In section \ref{sec:2dpheno} we discuss the phenomenology of these phases, and compare them to Fermi liquids. Section \ref{sec:threed} discusses how the results of the previous sections generalize to 3+1D. 

In section \ref{sec:mnsi} we consider an application of the general theory put forth in previous subsections. We consider systems consisting of a Fermi liquid coupled to a Bose-Luttinger liquid, and examine the effect that this coupling has on the transport properties of the Fermi liquid. A concrete example of a material where such a theoretical description may be applicable is the helical magnet MnSi, which exhibits a metallic phase possessing spin fluctuations whose dispersion has a degenerate minimum along a sphere in momentum space. Modeling this system as a Fermi liquid coupled to a Bose-Luttinger liquid, we calculate the transport scattering rate and show that it predicts a resistivity scaling as $\r \propto T^\eta$, where $\eta>1$ is a non-universal exponent. This offers a possible explanation for the observed $T^{3/2}$ scaling of the resistivity in this material,\cite{doiron2003fermi} which cannot be explained within the context of Fermi liquid theory alone.  

We close with a discussion of future lines of work in section \ref{sec:disc}, with discussions of a related model lacking $U(1)$ symmetry and several technical details relegated to the appendices.

The idea of using unconventional dispersion relations to stabilize higher-dimensional Luttinger liquid-like states has in fact already appeared in an earlier work by Sur and Yang,\cite{sur2019metallic} who focused on the context of spin-orbit coupled bosons in 2+1D.\footnote{We thank Zhen Bi for bringing Ref. \onlinecite{sur2019metallic} to our attention.} While the general idea of Ref. \onlinecite{sur2019metallic} is quite similar to that of the present paper, there are several key differences. Similar to  Ref. \onlinecite{sur2019metallic} we analyze the IR theory by decomposing the Bose surface into a large number of coupled Luttinger liquids. Unlike in Ref. \onlinecite{sur2019metallic} however, we take care to ensure that the physical properties of the IR theory do not depend on the exact way we perform this decomposition, which leads to a more careful analysis being needed when considering theories defined at finite density.  We also emphasize the importance of gapped vortex excitations which do not seem to have been considered in Ref. \onlinecite{sur2019metallic}. Our identification of the emergent symmetry is also different, and this leads us to  a different perspective on certain vortex operators. We argue that our treatment is needed in order to be confident about the stability of the theories we study. Finally, the present work is also slightly broader in scope, and includes discussions of several other related models, a procedure for performing RG, and an expanded treatment of various phenomenological aspects. 

\section{Warmup: 1+1D \label{sec:oned}}

In this section we will warm up by looking at the case of translation-invariant conserved bosons in 1+1D. We will be working at $T=0$ throughout, and will assume the presence of a reflection or time reversal symmetry ensuring that the dispersion is symmetric under $\bfk \ra -\bfk$. For simplicity will furthermore assume the existence of a particle-hole symmetry $P$ which fixes the average density of the bosons to be zero. This symmetry is imposed purely for simplicity, and all of the results in this section can easily be extended to the finite density case. 

The Bose-Luttinger liquids we will find in 1+1D are nothing more than multi-component Luttinger liquids endowed with a certain symmetry action. In 1+1D a Bose surface just consists of two points, and so these do not really give us examples of phases with a ``large'' number of low energy degrees of freedom. Nevertheless the analysis here is quite simple, and will be useful when we proceed to the more complicated 2+1D case.  

\ss{UV theory} 

\begin{figure}
	\centering 	
	\includegraphics{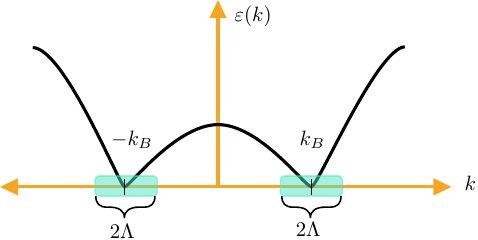}
	\caption{\label{1ddispfig} An illustration of a dispersion possessing minima at the two Bose points $\pm k_B$. The IR theory contains only modes within $\pm \L$ of each Bose point. }	
\end{figure}

Our starting point will be a Lagrangian whose free part gives a dispersion $\ep(k)$ possessing two minima at $\pm k_B$, with $k_B > 0$. The prototypical example of a Lagrangian with such a dispersion is 
\be \label{1dphl} \mcl = \psi^*\(-v\inv\p_\tau^2 + \frac{v}{4k_B^2}(-\p_x^2 - k_B^2)^2 + r \)\psi + \frac g4 |\psi|^4.\ee

We will be interested in the regime where $r<0$, so that the system is nearly a superfluid. The particle-hole symmetry acts as $P : \psi \mt \psi^*$, and the dispersion $\ep(k)$ is illustrated in Figure \ref{1ddispfig}.
One example of a system that exhibits this type of dispersion is the lower band of a spin-orbit-coupled boson,\cite{po2015two} although in what follows we will not restrict our attention any particular physical realization. 

To understand the IR theory we will integrate out modes at momenta far away from $\pm k_B$, assuming that the interaction $g$ is initially weak in the UV. After integrating out these modes, we obtain an effective action for the modes with momenta within $\pm \L$ of $\pm k_B$, where $\L / k_B \ll 1$ (see Figure \ref{1ddispfig}).

The two-body interaction of the bosons $g|\psi|^4$ is relevant under the free fixed point scaling, with RG eigenvalue $+2$. 
Since the flow is towards strong coupling, we will need to switch to a different language to describe the IR physics.

\ss{IR theory} 

Since we are assuming the interaction is weak in the UV, the kinetic energy is the dominant consideration when determining the correct IR Lagrangian to write down. We thus start by decomposing $\psi$ as 
\be \label{lrdecomp} \psi(x) = \frac1{\sqrt2}(e^{ik_Bx} \psi_R(x) + e^{-ik_Bx}\psi_L(x)),\ee  
with the $\psi_\lr$ fields which annihilate bosons at the right and left ``Bose points'' $k=\pm k_B$.
The symmetries of translation through a distance $a$ and $U(1)$ particle number act on the $\psi_\a$ fields as 
\be \label{psiasymmetries} U(1) \,:\, \psi_\a \ra e^{i\lambda} \psi_\a,\qquad T_a \,: \,\psi_\a(x) \ra e^{\a i k_Ba}\psi_\a(x+a),\ee 
where $\a = \pm1$ when it does not appear as a field index. 

In terms of these fields, the IR Lagrangian is  
\be \mcl = \sum_{\alpha=R,L} \psi_\a^*\( -v\inv \p_\tau^2 - v \p_x^2 + r\)\psi_\a + \sum_{\alpha\beta}g_{\alpha\beta} \psi^*_\a\psi_\a \psi^*_\b\psi_\b,\ee   
where $g_{\a\b}$ is a symmetric non-degenerate matrix parametrizing the interactions. 

In using the decomposition \eqref{lrdecomp} and in writing down the above Lagrangian, we have glossed over an important subtlety. Due to interactions the field $\psi$ will acquire a non-zero self-energy, which will generically renormalize the value of $k_B$. If this process is significant enough to renormalize $k_B$ to zero by the time we reach the IR scaling regime, a description in terms of the $\psi_\a$ fields will not be correct. In Appendix \ref{app:selfenergy} we argue that one can always choose the density and UV interaction strength such that the renormalized $k_B$ is finite, and henceforth we will always assume that this is the case. In the following, $k_B>0$ will then be taken to mean the renormalized Bose momentum. 

Since we are working at $r<0$, we are prompted to write $\psi_\a$ in terms of fluctuations about a nearly-superfluid state by taking 
\be \psi_\a = (r_0 + r_\a)e^{i\phi_\a},\ee 
where $r_0 = \sqrt{\r_S}$ is the square root of the average boson amplitude\footnote{We use the term ``boson amplitude'' here because there is no condensate ($|\lan \psi \ran | = 0$) and because ``superfluid density'' is potentially confusing, given that we are working at zero average boson density.} (since the action is $L\lra R$ symmetric, the potential favors an equal amplitude for both fields). The IR regime is reached at length scales larger than the inverse mass of the $r_\a$ fields. In this regime we may write down an IR Lagrangian solely in terms of the phase variables $\phi_\a$, which we take to have momentum modes in the interval $[-\L,\L]$.\footnote{Fixing a momentum cutoff of $\L$ on $\psi_\a$ is of course not the same as putting a cutoff of $\L$ on $\phi_\a$. A slightly more accurate treatment would be to use a sharp cutoff for $\phi_\a$ while using a soft cutoff for $\psi_\a$. Given that the exact cutoff procedure is not important for the effective field theory approach we are taking here, we will not pay attention to such subtleties in the following.} Fluctuations of $r_\a$ are accordingly taken into account by examining the effects of the vertex operators $e^{i\t_\a}$. Here $\t_\a$ are the fields dual to $\phi_\a$, with the commutation relations 
\be [\phi_\a(x),\p_x \t_\b(y)] = \a \twp i \, \d_{\a,\b} \d(x-y).\ee 
From \eqref{psiasymmetries} we see that $\phi_\a$ transforms under the relevant microscopic symmetries as 
\bea  U(1) \, & :\, \phi_\a \mt \phi_\a + \l, \\
T_a \, & : \, \phi_\a \mt \phi_\a + \a a k_B, \\ 
P \, & : \, \phi_\a \mt - \phi_{-\a},\eea 
with $\l \in [0,\twp)$ a constant and where $-\a$ denotes the opposite index to $\a$. 
The dual fields $\t_\a$ are neutral under $U(1)$ and $T_a$, and transform as $P : \t_\lr \mt \t_\rl + \pi$ under particle-hole symmetry. 

These considerations then lead to an IR Lagrangian which generically takes the form $\mcl_0+\mcl_I$, with
\bea \label{ir_lagrangian_oned} \mcl_0 & = \frac{1}{\fpi \eta} \sum_{\a} \( v\inv (\p_\tau\phi_\a)^2 + v(\p_x \phi_\a)^2\) \\
& \qquad \qquad + \frac{1}{\fpi \eta} \(v\inv f_\r \p_\tau \phi_L \p_\tau\phi_R + v f_j \p_x\phi_L \p_x\phi_R\)  \\ \mcl_I & = g\sum_\a \cos(2\t_\a) + g_{LR}^\pm \cos(\t_L \pm \t_R) + \cdots, \eea 
where $\cdots$ are higher-derivative interactions and less relevant cosines (note that there are no symmetry-invariant cosines in the $\phi_\a$ variables). 
The parameter $\eta$ is a non-universal phenomenological coefficient, and the $f_j, f_\r$ are ``Landau parameters'' characterizing the couplings of the spatial current densities $(f_j)$ and the couplings of the charge densities $(f_\r)$. 
Positivity of the Hamiltonian requires $|f_\r|,|f_j|<1$.  

The Lagrangian is diagonalized using the fields 
\be \phi_\pm\equiv \frac{\phi_R\pm \phi_L}{\sqrt2},\qquad \t_\pm \equiv \frac{\t_R \pm \t_L}{\sqrt2}, \ee 
which have commutation relations 
\be [\phi_\pm,\p_x\t_\pm]=0,\qquad  [\phi_\pm,\p_x\t_\mp] = \pm \twp i \d(x-y).\ee  
In terms of these variables, the Lagrangian is 
\be \label{tpmaction} \mcl_0 = \frac1\fpi  \sum_{\s=\pm} \frac1{\eta_\s} \phi_\s (v_\s\inv \p_\tau^2 + v_\s \p_x^2) \phi_\s,\ee 
where
\be \eta_\pm \equiv  \frac{\eta}{\sqrt{(1\pm f_\r)(1\pm f_j)}},\qquad v_\pm \equiv v \sqrt{\frac{1\pm f_j}{1\pm f_\r}}.\ee 
By dualizing the Lagrangian $\mcl_0$ in terms of the $\t_\pm$ variables (under which $\eta_\pm \ra 1/\eta_\pm$), one finds that the RG eigenvalues $y_\mco = 2 - \De_\mco$ of the most relevant interactions in $\mcl_I$ are 
\be y_{\cos(\t_\a)} = 2-\frac{\eta_+\inv + \eta_-\inv}4,\qquad y_{\cos(\t_L\pm \t_R)} =2- \frac1{\eta_\mp}.\ee 
If any of these eigenvalues are positive, some or all of the low-energy degrees of freedom will be made massive. 
However, it is always possible to choose $\eta$ small enough such that all three of the RG eigenvalues above are negative, and as such there always exists a regime of parameter space where the free fixed point described by $\mcl_0$ is stable. 

The phenomenology of the fixed point $\mcl_0$ can be determined straightforwardly, since the IR theory is simply that of two coupled Luttinger liquids acted on by translation and $U(1)$ symmetries in a particular way. Correlation functions at the fixed point are characterized by the non-universal exponents $\eta_\pm$, and possess oscillations at wavevectors corresponding to integer multiples of $k_B$. For example, the 2-point function of the UV bosons is 
\bea \label{1dpsicorr} \lan \psi(x) \psi^\da(0)\ran 
&\sim \frac{\cos(k_Bx)}{|x|^{\ob\eta}},\qquad \ob\eta\equiv \frac{\eta_+ + \eta_-}{2}.\eea
Rather than pursuing a detailed analysis of the phenomenology at this fixed point we will instead proceed directly to 2+1D generalizations, which is where our main interest lies.

\section{2+1D: UV theory and patch decomposition \label{sec:twod}} 

We will now turn our attention to systems of translation-invariant conserved bosons in 2+1D. As in the previous section, we will assume the presence of a UV particle-hole symmetry, which fixes the average particle density at zero and forbids a linear time derivative $\psi^* i\p_t \psi$ from appearing in the action. In Section \ref{sec:finite_density} we will explain what happens when this symmetry is absent.
We will furthermore assume that the bosons have a dispersion with a minimum along a circle of radius $k_B>0$ in momentum space. In order that this degeneracy be exact, we will assume the presence of continuous rotational symmetry, although we will see later that this assumption is not essential, as long as the rotation-breaking terms are small. 

A general UV Lagrangian satisfying these criteria can be written as  
\bea  
\mcl & = |\p_\tau \psi|^2 + A|\D \psi|^2 + B|\D^2 \psi|^4 + r |\psi|^2 + \frac{g}{4} |\psi|^4,\eea 
where $A<0$. We will be interested in the regime where $r<0$, so that a superfluid-like description can be used in the IR. Such a scenario can arise in the context of FFLO-type superconductivity (with the field $\psi$ representing Cooper pairs) or in certain types of frustrated magnets (which will be discussed further in section \ref{sec:mnsi}), but in what follows we will not be concerned with any particular physical realization. We will find it convenient to parametrize the kinetic part of $\mcl$ in terms of the momentum $k_B$ minimizing the dispersion as  
 \bea \label{freeCircle} 
\mcl & = \psi^*\(-\p_\tau^2  + \frac{1}{4k_B^2}(-\D^2 - k_B^2)^2 + r\)\psi + \frac{g}{4} |\psi|^4.\eea

To obtain the IR theory, we first integrate out modes with large $|\bfk | -k_B$, producing a theory with modes supported on a momentum-space annulus of width $2\L \ll k_B$ surrounding the Bose surface.
As in the 1+1D case the renormalization of $k_B$ as the modes away from the Bose surface are integrated out will be finite, and one needs to worry about whether or not $k_B$ can in fact be renormalized to zero. We again argue in Appendix \ref{app:selfenergy} that one can choose parameters such that this is generically not the case, and in what follows we will use $k_B$ to denote the renormalized Bose momentum, which we assume to be nonzero. 
At energy scales much less than $k_B$, the dispersion will cause the low-energy fields to fluctuate in a quasi-1+1D fashion, giving rise to a theory which in the IR has the potential to be treated using an approach similar to the one used in the previous section.

\begin{figure}
	\centering 
	\includegraphics{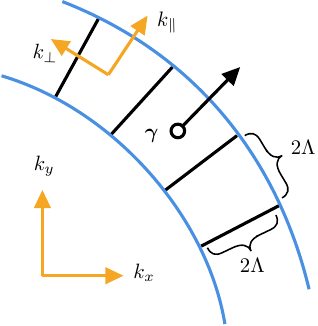}
	\caption{\label{FS_fig} How the low-energy annulus in momentum space is broken up into patches. Each patch is labeled by an angle $\g$, with corresponding unit vector $\bfga$.}
\end{figure}

After integrating out the modes far from the Bose surface it is helpful to use a patch decomposition for the remaining fields, similar to the ones employed in treatments of Fermi liquids.\cite{neto1994bosonization,houghton2000multidimensional,froehlich1997bosonization,shankar1994renormalization,sur2019metallic}
We proceed by breaking up the annulus around the Bose surface into many small patches of size $2\L \times 2\L$, and define patch fields $\psi_\g$ such that 
\be \label{patch_decomp} \psi(\bfx) = \frac1{\sqrt N}\sum_\gamma e^{ik_B \bfga \cdot \bfx} \psi_\gamma(\bfx),\ee 
where the momentum modes of $\psi_\g$ lie within a patch centered on the momentum $k_B \bfga$, with $\bfga$ a unit vector (see Fig. \ref{FS_fig}). The parameter $N$ is defined as the number of patches, viz.  
\be N \equiv  \frac{\twp k_B}{2\lap} \gg 1. \ee 
The kinetic term for each $\psi_\g$ field has the form 
\bea \label{expandedDispersion} \mcl_0 & \supset  \frac1{N} \psi^*_\g \Big( k_\prl^2 + \frac{1}{k_B}(k_\prl^3 + k_\prl k_\perp^2) \\ & \qquad\qquad   + \frac1{4k_B^2}(k_\prl^4 + 2k_\prl^2 k_\perp^2 + k_\perp^4)\Big)\psi_\g,\eea 
where $k_\prl = \bfk \cdot \bfga$ is parallel to $\bfga$, $k_\perp = \bfk \cdot \bfga_\perp$ is perpendicular, and where the notation $a \supset b$ is to be read as ``$b$ is a term appearing in $a$''. Note that due to the flatness of the dispersion along the $\bfga_\perp$ direction, there is no quadratic term $k_\perp^2$ appearing in the above kinetic term. 
Since $k_\prl,k_\perp$ are always much less than $k_B$, for most purposes we may approximate this as 
\be \label{lineardisp} \mcl_0 \supset \frac1{N}\psi_\g^* k_\prl^2 \psi_\g .\ee  
For some calculations it is however important to retain all the terms in \eqref{expandedDispersion}, as we will see when we discuss long-distance real-space correlation functions in section \ref{sec:2dpheno}. Until then, we will simply take the dispersion for each patch field $\psi_\g$ to be given by \eqref{lineardisp}. 

As a brief aside, we note that the exact procedure we use for breaking up the region near the Bose surface into patches is rather arbitrary, and should not have any bearing on the universal aspects of the IR theory. 
In particular, no physical quantities should have any explicit dependence on $N$ (indeed, we will see that $N$ flows under RG), which is something we will need to check as we go forward.

In terms of the $\psi_\g$ fields, the Lagrangian can be written as
\bea \label{hgamma} \mcl & = \frac1N\sum_\gamma  \psi^*_\g\(-  \p_\tau^2 - \D_\gamma^2 + r\) \psi_\g + \mcl_I,\eea
where $\mcl_I$ contains the interactions and where we have used the notation $\D_\g \equiv \bfga \cdot \D$.

As in Fermi liquids, the kinematics of the Bose surface ensures that the dominant interactions only occur in the forward-scattering and BCS channels, so that $\mcl_I = \mcl_{FS} + \mcl_{BCS}$, with 
\bea \label{mcli_psirep} \mcl_{FS} & = \frac1{4N^2} \sum_{\g,\gp} \psi^*_\g \psi_\g g_{FS}(\g-\gp)  \psi^*_\gp \psi_\gp \\ 
\mcl_{BCS} & = \frac1{4N^2} \sum_{\g,\gp} \psi^*_\g \psi^*_{\g+\pi} g_{BCS}(\g-\gp)  \psi_\gp \psi_{\gp + \pi},\eea 
where due to rotational symmetry the two interactions are functions only of angular differences. We now turn to writing down a Lagrangian which captures the IR physics of this theory.

\section{IR theory \label{sec:2dirtheory}} 

We flow to the IR by integrating out modes of $\psi_\g$ with large $(\bfk \cdot \bfga)^2 + \o^2$. 
As in 1+1D, the relevance of the density-density interactions forces us to switch to a different description for discussing the IR physics.
Since we are taking $r<0$ in \eqref{freeCircle}, we are prompted to minimize the potential $r|\psi|^2 + \frac g4 |\psi|^4$ by writing each patch field as
\be \psi_\g = (r_0+r_\g) e^{i\phi_\g}.\ee 

In what follows we will make the crucial assumption that the potential for the $\psi_\g$ fields favors a state where the expectation value $\lan r_0 + r_\g\ran$ is nonzero and independent of $\g$.\footnote{Allowing the expectation to be nonzero but with nontrivial $\g$ dependence is also possible, but we will ignore this possibility for now.} Depending on the details of the interactions in the UV this very well may not be the case, with the system preferring instead to spontaneously break rotation symmetry and develop amplitude order only at isolated points along the Bose circle.
Spontaneous symmetry breaking is energetically favorable if the UV interaction is a simple delta function contact interaction,\cite{brazovskii1975phase,pisarski2020transverse} 
although if the interaction acquires momentum dependence this need not be true.\cite{binz2006theory} 
There seems to be nothing a priori forbidding a state with uniform 
amplitude ordering for all of the $\psi_\g$,
and in what follows we will simply assume that this is the case.

Making this assumption, and working at length scales larger than the inverse mass of the $r_\g$ fields, we are lead to a superfluid-like IR description in terms of the phase fields $\phi_\g$. These fields are acted on by the microscopic $U(1)$ symmetry as 
\be \label{iru1action} U(1) \, :\, \phi_\g \mt \phi_\g + c \ee 
for constant $c$, while translation along a vector $\bfmu$ acts via 
\be \label{irtransaction} T_\bfmu \, : \, \phi_\g(\bfx) \mt \phi_\g(\bfx+\bfmu) + k_B\bfmu\cdot\bfga.\ee 
Finally, particle-hole symmetry sends $P : \phi_\g \mt - \phi_{\g+\pi}$. 

The general IR Lagrangian consistent with these symmetries is $\mcl_0+ \mcl_f + \mcl_I$, with $\mcl_I$ containing interactions (which will be discussed shortly in section \ref{sec:interactions}) and with the first two terms given by 
\bea \label{iraction} \mcl_0 & = \frac{k_B}{\fpi N \eta} \sum_{\g} \(v\inv (\p_\tau \phi_\g)^2  + v (\D_\g \phi_\g)^2\) \\ 
\mcl_f & = \frac{k_B}{\fpi N^2 \eta} \sum_{\g,\gp} \Big(v\inv f_\r^{\g,\gp} \p_\tau \phi_\g\p_\tau \phi_\gp   \\ & \qquad \qquad + v f_j^{\g,\gp}\D_\g \phi_\g \D_\gp \phi_\gp \Big),\eea 
where $\eta,f^{\g,\gp}_\r,f^{\g,\gp}_j$ are all dimensionless non-universal parameters, and where $v$ is a non-universal velocity. 
Fluctuations in the charge (current) density of the $\psi_\g$ fields at each patch are represented in the IR as $\p_\tau \phi_\g$ (as $\D_\g \phi_\g$), with \eqref{iraction} consequently being a general hydrodynamic Lagrangian parametrizing the gradient energy for fluctuations in the densities and currents.
The theory described by this fixed point is a Bose-Luttinger liquid (BLL), and is the fixed point that we will focus on for the majority of the rest of this paper.  
 
As in 1+1D, the couplings $f_j^{\g,\gp}$, $f^{\g,\gp}_\r$ are dimensionless ``Landau parameters'' characterizing the IR theory.
The $f_j^{\g,\gp}$ term couples the spatial current densities for the $U(1)$ particle number symmetries on patches $\g,\gp$, while $f_\r^{\g,\gp}$ couples the charge densities.
Due to rotational invariance, the Landau parameters will be functions only of $\g-\gp$. We will see that they are marginal under RG, just as in a Fermi liquid. 

While in some respects the BLL of \eqref{iraction} is similar to a bosonic Fermi liquid, there are several key differences. First, 
bosonized descriptions of Fermi liquids have only one pair of fields $\phi,\t$ for every pair of antipodal points, which is half as many as in the present context. 
Secondly, the coefficient $\eta$ (which we will see determines scaling of correlation functions at the fixed point) can take on any value, and is a non-universal function of the microscopic parameters.\footnote{In CFT language, $\eta$ is related to the radius of the $\phi_\g$ bosons as $\eta = 1/R^2$.} This is in contrast to the Fermi liquid context, where the value of $\eta$ is fixed.
Finally, in a Fermi liquid the spatial and temporal components of the current are related to one another by the Fermi velocity, and thus Fermi liquids have only a single set of Landau parameters. Here however the charge and current densities are distinct, giving two distinct sets of Landau parameters.

\ss{Emergent symmetry group} 

As in a Fermi liquid, the BLL fixed point possesses a very large emergent symmetry group. As formulated in \eqref{iraction} this symmetry group is naively realized by shifting 
\be \label{large_symmetry} \phi_\g(\bfx) \mt \phi_\g(\bfx) + f_\g(\bfx \cdot \bfga_\perp),\ee 
for some functions $f_\g$ of the perpendicular coordinate $\bfx\cdot\bfga_\perp$. 
This symmetry group is much too large however, and is an artifact of approximating the dispersion in each patch $\g$ by a function only of $\bfk \cdot \bfga$.\footnote{As mentioned earlier, this approximation does not change the analysis of the stability of the fixed point (to be discussed shortly), but is in fact too crude of an approximation for analyzing several physical properties of the fixed point. Therefore the fixed point theory technically must still remember the curvature of each patch, and the transformations \eqref{large_symmetry} cannot actually be the right emergent symmetry at the fixed point. } Accounting for the small curvature in each patch reduces the symmetry action \eqref{large_symmetry} to $\g$-dependent constant shifts. Since each $\phi_\g$ is a phase variable, the naive emergent symmetry group is then $U(1)^N$.

This is not correct though, as the way of tiling the region near the Bose surface into patches is arbitrary. While using square patches of size $2\L\times 2\L$ is a particularly convenient choice, we could equally well consider a decomposition into a larger number of narrower patches. Since the physical emergent symmetry group at the fixed point cannot depend on an arbitrary choice like this, identifying the emergent symmetry as $U(1)^N$ is clearly not correct. 

One might then think that since we are interesting in the large $N$ limit, we should simply identify the emergent symmetry group with $U(1)^\infty$.\cite{shankar1994renormalization,haldane2005luttinger,sur2019metallic} This is also not correct. A symmetry group of $U(1)^\infty$ would imply that the particle number at each point of the Bose surface is quantized to be an integer, but in fact we may only speak of a non-quantized particle density $\rho_\g d\g$, with the only quantized charge being the global charge $\int \frac{d\g}\twp \r_\g$.
 Furthermore, elements in $U(1)^\infty$ generically shift the $\phi_\g$ fields by discontinuous functions of $\g$. When we weakly break this symmetry group by e.g. adding a very small magnetic field (which introduces derivatives $\p/\p\g$ into the action), these discontinuous shifts create field configurations with infinite action, which we regard as unphysical. 

The correct identification of the emergent symmetry group is instead the loop group $LU(1) = {\rm Map}[S^1 \ra S^1]$.\cite{else2020non} This group acts on the $\phi_\g$ fields as 
\be \label{lu1action} LU(1) \, : \, \phi_\g \mt \phi_\g + f_s(\g),\ee 
where $f_s(\g)$ is a {\it smooth} function of $\g$, with $f_s(2\pi) - f_s(0) \in \twp \zz$ (the UV $U(1)$ particle number symmetry is embedded as the subgroup where $f_s$ is independent of $\g$, which is in fact the only $U(1)$ subgroup of $LU(1)$). The emergent symmetry group of $LU(1)$ is shared by the ``Ersatz Fermi liquids'' of Ref. \onlinecite{else2020non}.

Another way to arrive at this conclusion is to declare that only field configurations $\phi_\g$ which are smooth functions of $\g$ are physical, as this subspace is only preserved by $LU(1)$, and is violated by generic elements in $U(1)^\infty$. Our statement above about charge quantization can then be understood by noting that although the vertex operators $e^{in\phi_\g}$ are only well-defined for $n\in \zz$, it is not correct to treat the $\phi_\g$ as independent compact variables, since shifting a single $\phi_\g$ by $\twp$ cannot be done while obeying the smoothness constraint. Since each $\phi_\g$ is not individually compact, the charge on each patch is not quantized. The only compact variable is instead $\int \frac{d\g}\twp \, \phi_\g$, whose compactness ensures the quantization of the UV $U(1)$ charge. 

The $LU(1)$ symmetry is unfortunately not completely manifest in our description \eqref{iraction} of the fixed point, and is only made explicit if we sub-divide each square patch $\g$ into many infinitesimally thin slices. As already discussed, the price of doing this is that writing down Lagrangians which are local in real space becomes rather unwieldy. Therefore in what follows we will continue to work with the a finite number of square patches, with the acknowledgment that true emergent symmetry group is in fact $LU(1)$, and not $U(1)^N$.

Ref. \onlinecite{else2020non} showed that a large class of translation-invariant compressible (definable over a continuous range of densities) systems must necessarily have an infinite-dimensional emergent symmetry group in the IR, with $LU(1)$ being the simplest example. Despite the fact that the discussion above has been focused on the particle-hole symmetric zero-density limit, the BLL fixed point considered here in fact represents a compressible phase of matter, as we show in section \ref{sec:finite_density}. Thus one may ask whether the existence of the emergent $LU(1)$ symmetry is a necessary feature of the IR theory. 

However, we can not actually directly use the results of Ref. \onlinecite{else2020non}, which assumes that the IR symmetry group does not include any continuous higher form symmetries, which are symmetries whose charged objects have dimension greater than zero.\cite{gaiotto2015generalized} This assumption is actually violated in the present context: the BLL fixed point possesses a continuous 1-form symmetry associated with the fact that the worldlines of vortices in the UV boson field $\psi$ must form closed loops. 
A vortex in $\psi$ causes a simultaneous vortex in every $\phi_\g$ field, and is well-defined due to the quantization of the global $U(1)$ charge. This global vortex is massive at the BLL fixed point, and does not show up in the IR description. Nevertheless it must be included so that the IR and UV theories live in the same Hilbert space, and the additional 1-form symmetry it leads to means that the $LU(1)$ symmetry is not a priori a necessary feature of BLL-like fixed points, at least within the context of the filling  constraints of Ref. \onlinecite{else2020non}. Simpler examples of compressible states of matter with emergent continuous one-form symmetries and their formal properties will be discussed in Ref. \onlinecite{else2020qlm}.

\ss{Allowed perturbations to the fixed point \label{sec:interactions}}

In order to assess the stability of the fixed point described by $\mcl_0 + \mcl_f$, we need to know the interactions that can appear in $\mcl_I$, which we treat as perturbations to the fixed point. 
Any allowed perturbation must respect the UV symmetries of translation and $U(1)$ charge conservation. The most relevant symmetry-allowed interaction of the $\phi_\g$ fields is the BCS pairing term
\bea  \label{mcli} \mcl_I & \supset  \frac{1}{N^2} \sum_{\gamma,\gamma'}  g_{BCS}(\gamma-\gamma')\cos(\vp_{\g,\g'}),
\eea 
 where we have defined 
\be \label{vphidef} 
\vp_{\g,\g'} \equiv \phi_\g + \phi_{\g+\pi} - \phi_\gp - \phi_{\gp+\pi}.\ee 
This coupling explicitly breaks the $LU(1)$ symmetry of the fixed point down to the subgroup generated by functions with odd angular momenta, under which $\vp_{\g,\gp}$ is invariant. 
Since we are working with spinless bosons, the BCS coupling must consist only of even angular momentum channels, with $g_{BCS}(\g-\g') = g_{BCS}(\g-\gp+\pi)$. 

One important question to ask is whether or not cosines of the fields $\t_\g$ dual to $\phi_\g$ may appear in $\mcl_I$. The most natural way of defining these fields is to have them satisfy the commutation relations 
\be \label{phipmcomm} [ \phi_\g(\bfx),\D_\gp \t_\gp(\bfy)] =\twp i\, \frac{N}{k_B}\,\d_{\g,\gp}  \d^{(2)}(\bfx-\bfy),\ee 
so that exponentials of $\t_\g$ create vortices in the phases of the $\psi_\g$ patch fields.
The $\t_\g$ are neutral under the microscopic $U(1)$ symmetry, and since we are working at zero density in this section they are invariant under translation as well. Thus from the basis of the symmetry actions alone, one may also think to include in $\mcl_I$ cosines like 
\be \mcl_I \stackrel{?}{\supset} \frac1N \sum_\g g_\t \cos(\t_\g). \ee

We claim however that cosines in the $\t_\g$ fields do {\it not} represent legal perturbations to the fixed point (unlike in Ref. \onlinecite{sur2019metallic}), and that we may in fact restrict our attention purely to the pairing term \eqref{mcli}. 
There are several ways to argue this,\footnote{For a related discussion in the context of Fermi liquids, see Ref. \onlinecite{mross2011decohering}.} with the arguments being similar to those used when discussing the correct identification of the emergent symmetry group. First, the existence of well-defined vortex operators $e^{i\t_\g}$ would require the charge on each patch to be quantized. As was discussed above this is not the case, and only the global charge $\int \frac{d\g}\twp \r_\g$ is quantized.
Furthermore, the action of any putative vortex operator $e^{i\t_\g}$ would create a field configuration which is singular as a function of $\g$, which would have infinite action in the presence of a small $LU(1)$-breaking perturbation like a small magnetic field. Since there is no way to smoothly pass between field configurations of different vorticity, it is impossible to define a ``smoothed-out'' version of $e^{i\t_\g}$ which creates allowable non-singular field configurations. For these reasons, we will regard individual vortex operators $e^{i\t_\g}$ as being unphysical. Thus the only allowed perturbation to the fixed point is indeed the pairing interaction of \eqref{mcli} (as well as less-relevant higher-body operators). 

While operators creating vortices in each of the $\phi_\g$ fields individually are not allowed, there is of course always an allowed operator which creates a vortex in the UV field $\psi$. 
These vortices will be gapped excitations of the BLL phase. The low energy description in terms of the phase fields that we have developed is only legitimate at energy scales below the vortex gap. Indeed, the phase-only theory of the BLL does not know about the periodicity of the phase of $\psi$, and we need to incorporate these gapped vortex excitations in order to have an IR  theory that lives in the correct microscopic Hilbert space that $\psi$ lives in. From a formal point of view, the IR theory of the BLL without the vortices has a $U(1)$ one-form symmetry which is not present in the UV theory, and therefore we must also include excitations which explicitly break this one-form symmetry. An effective action that includes both the gapless excitations and the vortex field can be written along the same lines as the discussion for 2+1D bosonized Fermi liquids in Ref. \onlinecite{mross2011decohering}, but we will not do so explicitly here. Despite the fact that the vortices do not appear in the IR theory, we will argue in section \ref{sec:finite_density} that they play a crucial role in understanding how the BLL can exist at a generic non-zero density.

\ss{Fixed-point correlation functions \label{sec:correln_funcs}} 

Before determining the relevance of the terms in $\mcl_I$, let us first calculate the correlation functions at the fixed point described by $\mcl_0$. 
When the Landau parameters vanish, the two-point functions of the $\phi_\g$ fields are obtained from the Lagrangian \eqref{iraction} as
\bea \label{phiphi_corr} \mcg^\phi_{\g,\gp}(\bfk,\o) & \equiv \lan \phi^*_\g(\bfk,\o) \phi_\gp(\bfk,\o)\ran  = \d_{\g,\gp}  \frac{\twp l_\L \eta}{\o^2/v + v k_\g^2} 
\eea 
where we have defined 
\be l_\L \equiv \( \int_{-\L}^\L \frac{dk_\perp}{\twp}\)\inv = \frac\pi\L = \frac{N}{k_B} \ee 
as the length scale on which the patch fields can be localized. 

The effects of the Landau parameters show up only at order $1/N$, and as such can be ignored for the purposes of computing the patch field correlators.  
For example, if we consider the simple case where $f^{\g,\gp}_\r = f_\r$ is independent of angle and $f_j^{\g,\gp}=0$, we can show that $f_\r$ modifies the $\phi_\g$ correlators as  
\bea \label{frhocorreln} & \mcg^\phi_{\g,\gp}(\bfk,\o) = \frac{\twp l_\L \eta }{w^2/v + vk_\g^2} \d_{\g,\gp}  \\ 
& \qquad  - \frac{\twp l_\L \eta }{N(\o^2 + v^2 k_\g^2)(\o^2 + v^2 k_\gp^2)} \frac{\o^2 f_\r}{1 + f_\r \o /\sqrt{\o^2 + v^2k^2}},\eea 
where the square root in the last term comes from an angular integral over the Bose surface. 
The fact that the Landau parameters only enter at order $1/N$ (provided they are smooth functions of $\g-\g'$) is true for essentially the same reason as the statement that non-singular Landau parameters cannot destroy the quasiparticle in Fermi liquids,\footnote{This is only true in spatial dimensions greater than 1. In 1+1D we have $N=2$, and as we saw the Landau parameters do contribute an order 1 term to the self energy.} with the fact that the leading contribution to the self energy goes as $1/N$ being a standard feature of large-$N$ theories (this is essentially equivalent to the fact that mean field theory becomes exact as $d\ra\infty$).

In Fermi liquids, this means that destroying the quasiparticle with interactions is difficult. In the present context we are similarly unable to use the Landau parameters to make an order 1 modification to the self energy, but since we are starting from Luttinger liquids of arbitrary radius on each patch, we are still able to construct a theory without quasiparticles, as we will see shortly. 

The above discussion by no means implies that the Landau parameters have no physical consequences (as they make nonzero contributions to correlation functions involving integrals over the Bose surface), and we will see that they play an important role in some aspects of the phenomenology of the BLL fixed point. We will however set both Landau parameters to zero until we discuss this phenomenology in section \ref{sec:2dpheno}.

We now calculate the correlation functions of the vertex operators $e^{i\phi_\g}$ at the $\mcl_0$ fixed point. We find
\bea \lan e^{i\phi_\g(x)} & e^{-i\phi_{\g'}(0)} \ran \\ & \sim  \d_{\g,\g'} \exp\( - \twp l_\L v \eta  \int \frac{d^2k\, d\o}{(\twp)^3} \frac{e^{i\bfk\cdot\bfx + i\o \tau} - 1}{\o^2 + v^2 k_\g^2}\),\eea 
where the momentum integral is taken over the region $[-\L,\L]^2$. 
The integral in the exponent is 
\begin{widetext}
\bea \label{propintegral}2\pi l_\L v \eta \int \frac{d^2k\, d\o}{(\twp)^3}  \frac{e^{i\bfk\cdot\bfx + i\o \tau} - 1}{\o^2 + v^2 k_\g^2}   = - l_\L \eta  \int_{-\lap}^{\lap} \frac{dk_\perp}{\twp} \( \ln\( \frac1L \sqrt{x_\prl^2 + \tau^2 v^2} + \frac1{\L L} \)e^{ik_\perp x_\perp} + \ln(\L L)\),\eea 
\end{widetext}
where $L$ is an IR cutoff and $x_\prl = \bfx \cdot \bfga, \, x_\perp = \bfx \cdot \bfga_\perp$. 

When the perpendicular displacement $x_\perp \ll \L\inv$ the integral over $k_\perp$ is trivial, and simply cancels the factor of $l_\L$. When $x_\perp \gg \lap\inv$ the first logarithm term on the RHS of \eqref{propintegral} vanishes, and when this happens the remaining $\ln (\L L)$ term is uncanceled and sends \eqref{propintegral} to $-\infty$. Therefore we approximate the vertex correlator as 
\bea \label{squarepatch_vertex_correlator}\lan e^{i\phi_\g(x)} e^{-i\phi_{\g'}(0)} \ran & \sim \d_{\g\g'}\d_\L(\bfx \cdot \bfga_\perp) \\  &\qquad  \times \frac{1}{(1 +(\L v\tau)^2 + (\L \bfx \cdot \bfga)^2)^{\eta/2}},\eea 
where we have defined the function 
\be \label{dperpdef} \delta_\L(x_\perp) \equiv\begin{cases} 1, \qquad & |x_\perp| \leq \L\inv \\ 0 ,\qquad & {\rm else}\end{cases}. \ee 

Before moving on, let us comment briefly on the range in which our 
derivation of the correlation function 
\eqref{squarepatch_vertex_correlator} is valid. To derive this 
correlator, we have ignored the terms in the dispersion \eqref{expandedDispersion} depending on $k_\perp$. If we re-introduce the $k_\perp^4 / 
k_B^2$ term, the integral over $k_\perp$ means that the integral in \eqref{propintegral} no longer diverges logarithmically at long 
distances $x_\prl \gg \L\inv N$ \cite{moat}. Thus strictly speaking, the 
$e^{i\phi_\g}$ vertex operators have power-law correlations only for distances 
$\L\inv \ll x \ll \L\inv N$. This is however an artifact of discretizing the Bose surface, and the power-law behavior 
persists at all distances $\L\inv \ll x$ in the limit $N\ra\infty$.

\section{RG and stability \label{sec:2drg}} 

We are now interested in studying the stability of the fixed point governed by the Lagrangian $\mcl_0$ in \eqref{iraction}. We will find it convenient to use an RG scheme which is slightly different from the usual Fermi liquid approach,\cite{polchinski1992effective,shankar1994renormalization} which avoids any non-uniform re-scalings of spacetime. More details on this RG scheme and an application to Fermi liquid phenomenology can be found in Ref. \onlinecite{lake2021fermi}.

To perform RG, we first write $\phi_\g(\bfk,\o) = \phi^>_\g(\bfk,\o)+ \phi^<_\g(\bfk,\o)$, where $\phi_\g^>(\bfk,\o)$ consists of modes satisfying
	\be s\Lapr  < \sqrt{(\bfk\cdot\bfga)^2 +  \o^2/v^2} < \Lapr,\ee 
	where 
	\be s = 1 - d\ln \L\ee 
is a number slightly less than 1. We then
integrate out the $\phi_\g^>$, obtaining an effective action for the $\phi^<_\g$. Because after the mode elimination the resulting patches are no longer square, we further re-partition the low-energy annulus into slightly smaller square patches of size $2s\L\times 2s\L$, thereby increasing the number of patches to $N/s$. 
Finally, we re-scale the UV $\psi_\g$ fields as 
\be \label{psi_rescaling} \psi_\g \mt \sqrt s \psi_\g,\ee 
which preserves the $1/N$ normalization in the patch decomposition of $\psi$ \eqref{patch_decomp}. 

The RG flow of the couplings in $\mcl_I$ is obtained by comparing 
the dimensionless couplings before and after the mode integration. 
To evaluate the relevance of perturbations to $\mcl_0$, we then need to know how to construct dimensionless parameters from the couplings appearing in $\mcl_I$.

 In conventional scenarios, one is interested in RG flows near a scale-invariant fixed point. In that case there is only one scale in the problem (namely the cutoff $\L$), and as such there is a unique way of defining dimensionless coupling constants. In the present context however there is another scale, namely $k_B$. The Bose momentum $k_B$ is a defining momentum scale of the theory, and does not change during mode elimination. 
This means that if we make a given coupling constant $g$ dimensionless using powers of both $k_B$ and $\L$, only the powers of $\L$ will determine the RG eigenvalue of $g$. 

To determine the flow of a given coupling constant $g$, we then need to figure out the correct way of using powers of $k_B$ and $\L$ to define a dimensionless coupling constant $\bar g$. Consider for example the Landau parameters $f_\r$ appearing in the free Lagrangian $\mcl_f$ of \eqref{iraction}. As it stands the $f_\r$ are dimensionless, and since no powers of $\L$ appear in its contribution to the action, it will be marginal under RG. However, we could equally well keep $f_\r$ dimensionless while replacing the $k_B$ appearing in \eqref{iraction} with $\L$. In this case we would naively conclude that the $f_\r$ are relevant under RG. How do we resolve this ambiguity? 

To see the answer, recall that $k_B$ and $\L$ are related by $N = \pi k_B / \L$. Thus different ways of making coupling constants dimensionless differ from one another by powers of $N$. The correct dimensionless couplings are then chosen in a way such that the dimensionless couplings always make at most order $N^0$ contributions in perturbation theory to correlation functions at the fixed point. If instead a dimensionless coupling always makes $N^{n<0}$ contributions to correlation functions it can be ignored, while if it can make $N^{n>0}$ contributions then a perturbative RG analysis is invalid in the first place. 

For example, it is easy to show that as in Fermi liquids, the Landau parameters only appear in correlation functions in the combinations $f,f/N,f/N^2,$ and so on. Thus the Landau parameters are dimensionless and can be taken to be of order 1 as they appear in \eqref{iraction}, and as such are indeed marginal (the scaling of the $1/N^2$ in the Landau parameter term is canceled by the multiplicative re-scaling of the $\psi_\g$ fields appearing in \eqref{psi_rescaling}). 

The $\g$-index structure of the BCS term is the same as that of the Landau parameters, and similarly appears only in the combinations $g_{BCS} /(N\L^3), g_{BCS} / (N^2\L^3)$, etc. Thus the correct dimensionless coupling for the BCS interaction is 
\be  
\bar g_{BCS}(\g-\gp) \equiv  \frac1{N \L^3} g_{BCS}(\g-\gp),\ee 
so that $\mcl_I$ can be written as 
\bea  \mcl_I & = \frac{k_B \L^2}{N^2} \sum_{\gamma,\gamma'}  \bar g_{BCS}(\gamma-\gamma') \cos(\vp_{\g,\g'}).
\eea 
Thus the relevance of the BCS term is determined by comparing the dimension of $\cos(\vp_{\g,\gp})$ to 2 and not to 3, the actual dimension of spacetime (this is true even though there exist correlation functions of $\cos(\vp_{\g,\gp})$ having power-law behavior along all three spacetime directions). 

With this in mind, let us now discuss how to integrate out the fast modes. To do this, we will need to know correlation functions of the fast field vertex operators $e^{i\phi^>_\g}$. These are
\bea \lan e^{i\phi^>_\g(0)}\ran  = \exp\( - \eta \frac{d\ln \L}{2}\) \approx s^{\eta/2}  \eea 
and 
\bea \label{fastmodecorr} \lan e^{i\phi^>_\g(\bfx,\tau)} e^{-i\phi^>_\gp(0)} \ran  & = \exp\Big( \eta \, d\ln \L \Big[ \d_{\g,\gp}\d_\L(\bfx \cdot \bfga_\perp)  \\ & \times  J_0[\L\sqrt{(\bfx\cdot\bfga)^2 + v^2\tau^2}] - 1\Big]\Big),\eea 
where we have used $l_\L \int \frac{dk}\twp e^{i q y} \approx \d_\L(y)$. 

We can now integrate out the fast modes in the usual manner. The lowest-order contribution in $\bbcs$ to the effective action for the slow modes is 
\bea S_{eff}  & \supset \frac{k_B \L^2}{N^2} \sum_{\g,\gp} \int d^3x\, \bbcs(\g-\gp) \lan \cos(\vp_{\g,\g'}^< + \vp_{\g,\g'}^>)\ran ,\eea 
where the expectation value is taken with the free action for the $\phi^>_\g$ fields, and where $\vp_{\g,\g'}^{>/<}$ is the fast / slow mode part of $\vp_{\g,\g'}$. Separating out the cosine and using $\lan \sin( \vp^{>}_{\g,\gp})\ran=0$, we have 
\bea S_{eff} & \supset \frac{k_B\L^2}{N^2}  \sum_{\g,\gp} \int d^3x \, \bbcs(\g-\gp) \cos(\vp^<_{\g,\gp}) \lan e^{i\vp^{>}_{\g,\gp}(0)}\ran \\ 
& = \frac{k_B(s\L)^2}{N^2} s^{2\eta-2}\sum_{\g,\gp} \int d^3x \, \bbcs(\g-\gp) \cos( \vp^{<}_{\g,\gp}),\eea 
with $s\L$ the cutoff for the slow fields. 
The new dimensionless coupling is then $s^{2\eta-2}\bbcs$, which determines the RG eigenvalue of $\bbcs$ to be 
\bea y_{\cos(\vp_{\g,\gp})} & = 2 - 2\eta.\eea 
Thus the pairing interaction will be irrelevant provided that
\be \label{stability_cond} \eta > 1.\ee 
Loop contributions can be worked out in a similar fashion using the propagators \eqref{fastmodecorr}; doing this one finds
\be \label{bcsbeta} \frac{d\bar g_{BCS}^{l}}{dt} = (2-2\eta)\bar g_{BCS}^{l} - C(\bar g_{BCS}^{l})^2,\ee 
where we have defined the harmonics $\bar g_{BCS}^{l} = \int d\g \, \cos(l\g) \bar g_{BCS}(\g)$, and where $C$ is a positive constant. Since we are working with spinless bosons we can restrict to even harmonics with $l\in 2\zz$ (as $\bbcs(\g-\g') = \bbcs(\g-\gp+\pi)$). 
The most important difference with respect to the case of Fermi liquids is that here the pairing interaction $\bbcs$ is generically {\it not} marginal at tree-level.

If the pairing term is irrelevant, the IR physics is simply that of the BLL fixed point \eqref{iraction}, which we will explore further in the next section. 
Consider on the other hand the case where the pairing terms are relevant. 
If there exist angular momentum channels with $\bbcs^l<0$ we expect spontaneous symmetry breaking to occur, with 
\be \lan  \phi^+_\g \ran =  l_* \g + c.\ee 
Here $c$ is a constant (coming from the global $U(1)$ symmetry), $l_*$ is the angular momentum with the most negative $\bbcs^{l}$, and we have defined  
\be \phi^\pm_\g \equiv \phi_\g \pm \phi_{\gp}.\ee 
In the symmetry-broken phase the $\phi^+_\g$ are all given expectation values, while the $\phi^-_\g$ are unaffected (since the $\phi^-_\g$ are neutral under the global $U(1)$, they can never be gapped out by pairing interactions). The resulting phase is thus a rather unconventional paired superfluid, possessing a Bose surface and described in the IR with the remaining fields $\phi^-_\g$. This produces essentially the same IR theory as that of a BLL arising from a system of real bosons, as we discuss in appendix \ref{app:real}. 

If all of the $\bbcs^l$ are positive,\footnote{Even if all the bare couplings are positive, negative couplings still have the potential to be generated by a bosonic version of the Kohn-Luttinger mechanism.\cite{kohn1965new} As in Fermi liquids these effects are however likely to be very small, and in any case are only expected to matter at rather large $l$.} 
we cannot find a symmetry breaking pattern for the $\phi^+_\g$ which minimizes the cosines in the pairing interaction. However, we see from the beta function \eqref{bcsbeta} that at least to quadratic order, the flow for positive couplings is in fact towards a nontrivial fixed point with $\bbcs^{l} = (2-2\eta)/C$. We defer an exploration of this interesting fixed point to future work. 

Summarizing, we see that regardless of the value of $\eta$, there are no relevant perturbations to the BLL fixed point which are able to completely gap out the Bose surface. To pass into a trivial gapped phase without explicitly breaking a symmetry, one may tune the parameter $\mu$ in the UV Lagrangian \eqref{freeCircle} to be negative, or modify the dispersion such that $k_B$ is taken to zero. One may presumably also pass to a Mott insulator by condensing the vortices for the UV $\psi$ bosons, although as mentioned earlier these vortices are massive at the fixed point and do not have a natural representation in terms of the IR fields. Figuring out how condense these vortices, as well as identifying the nature of the phase transition and resulting insulating state, are interesting questions that we leave to future work.

Finally, it is also important to also address the question of whether or not the BLL phase is stable with respect to small modifications of the UV dispersion. We have so far assumed a dispersion possessing rotational symmetry, but as we are ultimately interested in theories emerging from UV lattice models, this assumption will generically be violated. 

Consider then adding a small perturbation which breaks the continuous rotational symmetry of the dispersion down to some discrete subgroup, like $\d \ep \propto  k_x^4 + k_y^4$.
As long as the change in the dispersion caused by this perturbation is small compared to the energy scale at which the IR hydrodynamic description sets in, it can be dealt with by adding terms dependent on $\bfga_\perp \cdot \D$ to the dispersion for the $\phi_\g$ patch fields. The leading terms will be linear in $\bfga_\perp \cdot \D$, but since these become total derivatives in the $\phi_\g$ representation they can be ignored. More generally, since the correlation functions for $\phi_\g$ at the rotation-invariant fixed point do not depend on $\bfx \cdot \bfga_\perp$, the added terms dependent on $\bfga_\perp \cdot \D$ will not modify any of the fixed-point correlation functions within perturbation theory. Therefore the BLL phase is insensitive to rotation-breaking perturbations to the dispersion, provided they are small enough so that the fixed point Lagrangian \eqref{iraction} is still a good starting point for describing the IR theory.

\section{Generalization to finite density \label{sec:finite_density}}

Until now, we have been assuming the presence of a particle-hole symmetry which fixes the average particle density $\bar\rho$ to be zero.\footnote{Note that we are always distinguishing between the average particle density (viz. the expectation value of the generator of the $U(1)$ symmetry, whose form depends on the structure of the time derivative terms in the action) and the boson amplitude $\lan |\psi|^2\ran$. The boson amplitude is nonzero in all of the phases we consider, while the average particle density is nonzero only in the absence of particle-hole symmetry.} This limit is not required for stability of the BLL fixed point, and the BLL is in fact a compressible phase of matter, definable for a continuous range of densities. The generalization to the finite-density case requires some care however, which we now explain. 

Let us first look at the most obvious way of generalizing the discussion above to finite density, which was the approach taken in Ref. \cite{sur2019metallic} We start from the UV Lagrangian 
\bea \label{nonrelfreeCircle} 
\mcl & = \psi^*\(\p_\tau - \mu + \frac{1}{8mk_B^2}(-\D^2 - k_B^2)^2\)\psi + \frac{g}{4} |\psi|^4,\eea 
where the average density is fixed by $\mu>0$ and $g$. Note that we have not included a second order time derivative term $\psi^*\p_\tau^2\psi$, on the grounds that it is irrelevant under the non-relativistic $z=2$ scaling of the $g=0$ fixed point. 

Starting with this Lagrangian, we again decompose $\psi$ into patches, and make the assumption that each patch field is nearly a superfluid, so that we may write  
\be \psi_\g \ra \sqrt{\bar \r + \frac{ k_B}\twp \D_\g \t_\g} \,  e^{i\phi_\g},\ee 
where $\bar\r \neq 0$ is independent of $\g$ and where $k_B \D_\g\t_\g /\twp$ keeps track of long-wavelength fluctuations in the density on each patch. 
The hydro fields $\phi_\g,\t_\g$ are acted on by the microscopic $U(1)$ as 
\be \label{iru1action_nonrel} U(1) \, :\, \phi_\g \mt \phi_\g + c ,\qquad \t_\g \mt \t_\g,\ee 
for constant $c$, while translation along a vector $\bfmu$ acts via 
\bea \label{irtransaction_nonrel} T_\bfmu \, : \, & \phi_\g(\bfx) \mt \phi_\g(\bfx+\bfmu) + k_B\bfmu\cdot\bfga,\\  T_\bfmu \, : \, &  \t_\g(\bfx) \mt \t_\g(\bfx + \bfmu) + \frac{ \bar\r}{k_B}  \bfmu \cdot \bfga.   \eea 

Using this bosonized representation, the general hydrodynamic IR Lagrangian we are led to consider is then $\mcl_0+ \mcl_f + \mcl_I$, with $\mcl_I$ containing the BCS pairing interactions, and with the first two terms given by 
\bea \label{thetaphi_action_nonrel} \mcl_0 & = \frac1N \sum_{\g} \( \bar\r \p_\tau\phi_\g + \frac{k_B}\twp \D_\g \t_\g \p_\tau \phi_\g  + \b (\D_\g \phi_\g)^2 \) \\ 
\mcl_f & = \frac{1}{N^2} \sum_{\g,\gp} \Big( g_\t(\g-\g') \D_\g \t_\g  \D_\gp \t_\gp \\ & \qquad \qquad + g_\phi(\g-\gp) \D_\g \phi_\g \D_\gp \phi_\gp\Big),\eea 
where $\b\sim \bar\r / 2m$ and where the first term in $\mcl_0$ comes from the $\psi^* \p_\tau \psi$ term in \eqref{nonrelfreeCircle}. 

We then integrate out the $\t_\g$ fields, producing a term coupling the $\p_\tau \phi_\g$ on different patches. Doing this, we get 
\bea \label{iraction_nonrel} \mcl_0 & = \frac{k_B}{\fpi N \eta} \sum_{\g}  v (\D_\g \phi_\g)^2 \\ 
\mcl_f & = \frac{k_B}{\fpi N^2 \eta} \sum_{\g,\gp} \Big(v\inv f_\r^{\g,\gp} \p_\tau \phi_\g\p_\tau \phi_\gp  \\ & \qquad \qquad  + v f_j^{\g,\gp}\D_\g \phi_\g \D_\gp \phi_\gp \Big),\eea 
where $\eta,f^{\g,\gp}_\r,f^{\g,\gp}_j$ are again all dimensionless non-universal parameters. 

The most important difference between \eqref{iraction_nonrel} and the theory with particle-hole symmetry \eqref{iraction} is that here the only term producing stiffness for charge density fluctuations is the $f_\r^{\g,\g'}$ Landau parameter term arising from the density-density interactions of the $\psi_\g$ fields. The fact that there is no $(\p_\tau \phi_\g)^2$ term in the first line of \eqref{iraction_nonrel} is due to the absence of the $\psi^*\p_\tau^2\psi$ term in the UV Lagrangian, which provides a nonzero stiffness to the density fluctuations coming from the rest energy of the charges.
 In the absence of this term, there is nothing to provide an $O(N^0)$ stiffness for the charge fluctuations, since the Landau parameters only modify correlation functions of the $\phi_\g$ fields at order $1/N$. As a result, physical properties of the phase, including correlation function exponents, acquire explicit $N$-dependence. Unlike in Ref. \onlinecite{sur2019metallic}, our view here is that such dependence is unphysical (as $N$ flows under RG, for example), and as such we do not regard this approach as a route to obtaining a stable BLL phase.\footnote{In 1+1D this is not an issue, since there we have $N=2$, and the Landau parameters make a nonzero contribution to the correlation functions of the patch fields.}
 
Fortunately, we will now argue that the reasoning leading to \eqref{iraction_nonrel} is a bit too hasty. Indeed, we claim that instead of \eqref{nonrelfreeCircle}, the correct UV starting point is a Lagrangian containing a term with a quadratic time derivative, with 
\be \mcl \supset \psi^* \( \p_\tau - \mu - \frac{\l m}{k_B^2} \p_\tau^2 \) \psi,\ee
where $\l$ is a dimensionless parameter. 
While the $\l$ term is irrelevant under the $z=2$ UV scaling, in the IR variables $\phi_\g,\t_\g$, the $\l$ term in fact has the {\it same} scaling dimension as the linear $\p_\tau$ term (as it becomes $\sim N\inv \sum_\g (\p_\tau \phi_\g)^2$ in the IR representation), and therefore it should be kept. 

In particular, we will be interested in situations where the renormalized value of $\l$ is of order 1.
The amount of RG time required to reach the IR regime where $\l$ is marginal need not be very long, and depends on the exact values of the microscopic parameters (some further discussion can be found in appendix \ref{app:selfenergy}). Thus this assumption does not require any particular fine-tuning. 

With the $\l$ term, the IR theory includes an $O(N^0)$ patch-diagonal $(\p_\tau \phi_\g)^2$ term, giving the charge density fluctuations a finite stiffness as $N\ra\infty$. The IR theory at finite density thus takes on the same form as in the zero-density case (provided that the UV value of $\l$ is not too small), and as such the BLL phase is a compressible phase of matter. 

Finally, let us understand how the BLL reacts to a change in the average density $\bar\r$. As a compressible phase there is essentially no change in the gapless sector described by the phase modes $\phi_\g$. However, much like in the familiar superfluid phase, the gapped vortices will see the average particle density as an effective  background magnetic field. Thus translations will act projectively on these vortex degrees of freedom. As the particle density is changed the effective background magnetic field will change, and accordingly so to will the action of magnetic translations on the vortices. This is the main effect of changing the density, and is sufficient to ensure that the low energy theory has the correct action\cite{else2020non} of translation when a uniform $2\pi$ magnetic flux is turned on. See  Ref. \onlinecite{else2020qlm} for a discussion of these issues in a simpler context.

\section{Phenomenology \label{sec:2dpheno}} 

We now make some brief comments on the phenomenology of the BLL fixed point, assuming $\eta>1$ so that the pairing interactions in \eqref{mcli} are irrelevant. In some aspects the phenomoenology is similar to Fermi liquids, but in other aspects it is rather different. 

\ss{Thermodynamics} 

Since the IR theory is given by a collection of compact bosons with exactly marginal current-current interactions, the specific heat $C$ will always be linearly proportional to $T$, as in a Fermi liquid. To get an exact expression for $C$ we would need to diagonalize the Hamiltonian resulting from the Lagrangian \eqref{iraction}, which is nontrivial when the Landau parameters are nonzero. However, the Landau parameters only enter $C$ at order $1/N$, and as such can be ignored.
Since the specific heat (density) of a non-chiral 1+1D boson dispersing as $w = vk$ is $C_{1+1D} = \pi T / 3v$, we then have 
\be C = \frac{\pi T}{3 v} \frac{2 \pi k_B}{\twp} = \frac{\pi k_B T}{3v}.\ee 
Here $k_B$ should not be confused with Boltzmann's constant, which is set to unity throughout. 

The compressibility is calculated from the connected density-density correlator, the low-momentum part of which is 
\be \chi_{\r\r}(\bfk,\o) = 2\(\frac{k_B}{\fpi vN \eta}\)^2 \sum_{\g,\gp} \o^2 \lan \phi_\g(\bfk,\o)\phi_\gp(-\bfk,\o)\ran.\ee 
The compressibility is obtained from this correlation function by taking the limit $\o\ra0$ after setting $\bfk=0$. 
Since the current Landau parameters $f_j^{\g,\gp}$ do not contribute to correlation functions of the $\phi_\g$ fields at $\bfk=0$ the compressibility will not depend on them, and without loss of generality we can set them equal to zero. 

From the above we see that $\chi_{\r\r}$ is proportional to $\lan \phi_0(\bfk,\o) \phi_0(-\bfk,-\o)\ran$ where the charge mode $\phi_0$ is defined as 
\be \phi_0 \equiv \int \frac{d\g}{\twp}\, \phi_\g,\ee 
so that the compressibility is only sensitive to the zeroth Fourier mode $f^{(0)}_\r = \int \frac{d\g}{\twp} f^\g_\r$. Computing the correlation function with \eqref{frhocorreln}, we then find for the compressibility 
\bea \kappa & = 2N \( \frac{k_B }{\fpi vN \eta}\)^2 \twp l_\L v\eta\(1 - \frac{f_\r^{(0)}}{1+f_\r^{(0)}}\) \\ &= \frac{k_B}{\fpi v \eta } \frac1{1+f^{(0)}_\r},\eea 
which parametrically is the same as in a Fermi liquid, but with $k_F$ replaced by $k_B$.

\ss{Zero sound} 

Even though there is no quasiparticle having finite overlap with the UV boson $\psi$ (due to the continuous exponent appearing in the $e^{i\phi_\g}$ correlators), these theories can host collective zero sound modes in a manner similar to Fermi liquids. 
Charge and momentum are carried by separate fields, and as such we can consider collective modes in either the $\phi_\g$ phase variables or in the $\t_\g$ density variables. 

For example, consider the case where $f^{\g,\gp}_\r = f_\r, f_j^{\g,\gp} = f_j$ are both constants, so that the fixed-point Lagrangian reads (now in real time)
\bea \mcl & = \frac{k_B}{\fpi N\eta} \sum_{\g,\gp} \phi_\g \Big(\o^2v\inv(\d_{\g,\gp} + N\inv f_\r) \\ & \qquad \qquad  - k_\g k_\gp v(\d_{\g,\gp} + N\inv f_j)\Big)\phi_\gp.\eea 

The equation of motion for $\phi_\g$ reads 
\be  \phi_\g = \frac1{N(-\o^2  + v^2k_\g^2)}\sum_\gp (f_\r \o^2 - f_j k_\g k_\gp )\phi_\gp.\ee 
We now sum over $\g$, and replace $N\inv \sum_\g \ra \int \frac{d\g}\twp$. 
We see then that the $f_j$ term drops out, and that the equation of motion becomes 
\be \phi_0 = f_\r \o^2  \int \frac{d\g}{\twp} \frac1{-w^2 + v^2k^2\cos^2\g}\phi_0.\ee 
Nonzero solutions exist provided that $\o / vk > 1$ (so as to avoid the pole in the denominator), for which we can solve the above equation to find 
\be f_\r = -\sqrt{1-\(\frac{vk}{\o}\)^2},\ee 
in terms of which 
\be \o = \frac{vk}{\sqrt{1-f_\r^2}}.\ee 
Therefore zero sound modes arise at $\o > vk$ as long as $-1<f_\r<0$. Note that as in a Fermi liquid, the zero sound velocity is always greater than $v$. 

Collective modes of the dual $\t_\g$ fields are analyzed in a similar way. When we rewrite the free action in terms of the $\t_\g$ fields, we find 
\bea \mcl & = \frac{k_B\eta }{\fpi N} \sum_{\g,\gp} \t_\g \Big(\o^2v\inv(\d_{\g,\gp} + N\inv \wt f_\r) \\ & \qquad \qquad - k_\g k_\gp v(\d_{\g,\gp} + N\inv \wt f_j)\Big)\t_\gp,\eea
where the dual Landau parameters are
\be \wt f_\r = - \frac{ f_j}{1+f_j},\qquad \wt f_j = - \frac{f_\r}{1+f_\r},\ee 
which follows from $(\unit + a C)\inv = \unit - \frac{a}{1+a} C$, where $C$ is the $N\times N$ matrix with each entry equal to $1/N$.
Therefore using the same steps as above we conclude that regardless of $\wt f_j$, a collective mode in $\t_0(\o,k) \equiv \int \frac{d\g}{\twp} \t_\g(\o,k)$ exists provided that $\o/vk>1$ and $-1< \wt f_\r < 0$, with the dispersion being $\o = vk/\sqrt{1- \wt f_\r^2}$. Thus couplings of the $U(1)$ charge densities give rise to collective phase modes, while couplings of the $U(1)$ current densities give rise collective density modes.

\ss{Real-space correlation functions}

We now turn to studying the long-distance behavior of various correlation functions of the UV bosons $\psi$. When doing this, it is important to retain the subleading terms in the dispersion \eqref{expandedDispersion} in order to account for the fact that the Bose surface curves slightly within each patch. These effects show up on length scales larger than $ \sim k_B / \L^2$
and were not important when performing RG in the previous section, since the RG eigenvalues are calculated using the correlation functions of the fast fields at zero spacetime separation. When computing long-distance correlation functions however, the curvature within each patch must be accounted for. 

To do this, we refine each patch field $\psi_\g$ as 
\be \psi_\g(\bfx) = \int_{\g-\L/k_B}^{\g+\L/k_B} \frac{d\gp} \twp \, e^{ik_B(\bfga'-\bfga)\cdot \bfx} \, \wt\psi_\gp(\bfx),\ee 
with $\wt\psi_\g(\bfx)$ supported on an infinitesimally thin sliver of momentum space oriented along the $\bfga'$ direction. As we did for the $\psi_\g$ fields, we then continue to assume that we may work in a phase representation with $\wt \psi_\g \sim e^{i\wt\phi_\g}$. The free Lagrangian $\mcl_0$ is still diagonal when written in terms of the $\wt\phi_\g$,\footnote{Since $\wt\psi_\g$ is completely delocalized along $\bfx\cdot\bfga_\perp$, the $\wt\psi_\g$ fields are not well-suited for dealing with couplings between different angles on the Bose surface, which is why we did not make use of them above. These off-diagonal couplings however do not enter into the expression for the $\psi$ correlator, and so for the present purposes it is better to calculate with the $\wt\psi_\g$ fields.} and we find that the $e^{i\wt\phi_\g}$ have correlation functions
\be \lan e^{i\wt\phi_\g(x)} e^{-i\wt\phi_\gp(0)} \ran = \d_{\g,\gp} \frac1{(\tau^2 + (\bfx \cdot \bfga)^2 )^{\eta/2}},\ee 
with the only difference compared to the $e^{i\phi_\g}$ correlators being the complete independence on $\bfx \cdot \bfga_\perp$ (now and in the following, we will not be explicitly writing out the regularization by the UV cutoff or unimportant constant factors).
The correlation function for the $\phi_\g$ fields with the curvature in each patch taken into account is therefore  
\bea \label{thin_patch_vertex_correlator} \lan e^{i\phi_\g(x) } e^{-i\phi_\g(0)} \ran \sim \int_{\g-\L/k_B}^{\g+\L/k_B}   & \frac{d\g'}\twp \, e^{ik_B (\bfga' - \bfga)\cdot \bfx} \\ 
& \times  \frac1{(\tau^2 + (\bfx \cdot \bfga')^2)^{\eta/2}}.\eea

We now calculate the correlation functions of the UV bosons at long spacetime distances, $x\L \gg 1$. 
We find  
\bea \chi(\bfx,\tau) & = \lan \psi(\bfx,\tau) \psi^\da(0)\ran
\\ & \sim  \int \frac{d\g}\twp \, \lan e^{i\wt \phi_\g(\bfx,\tau)} e^{-i\wt \phi_\g(0)}\ran \\ & \sim 
\int \frac{d\g}{\twp}\frac{e^{ik_B x\cos\g }}{(\tau^2 + (x\, \cos\g)^2)^{\eta/2}}.\eea 

Consider now the case of purely spatial separation, with $\tau=0$. 
Since we are interested in $xk_B \gg x\L \gg 1$, only the angular regions near the stationary points of the exponential (viz. $\g = 0,\pi$) contribute significantly to the integral. Therefore we can ignore the $\cos\g$ in the denominator, with the integral over $\g$ then producing a term proportional to $J_0(k_B x \gg 1) \propto (k_Bx)^{-1/2} \cos(k_B x - \pi/4)$, and hence the leading contribution to $\chi(\bfx,\tau)$ takes the form 
\bea \label{uvbosoncorr} \chi(\bfx,0) & \sim \frac{\cos(k_B x - \pi / 4)}{x^{\eta+1/2}},\eea
which decays faster than any of the $e^{i\wt\phi_\g}$ by virtue of destructive interference from multiple patches. The phase shift of $\pi/4$ in \eqref{uvbosoncorr} is the same as one finds in Fermi liquids; unlike in Fermi liquids however, the exponent of the power law in \eqref{uvbosoncorr} is continuously tunable. 

Using the Fourier transformation of the patch vertex operators \eqref{thin_patch_vertex_correlator}, we see that the equal-time momentum-space expectation value of the $\psi$ fields is (recall that $\eta > 1$ for stability)
\bea \lan \psi_\bfk \psi^\da_\bfk\ran \sim |k-k_B|^{\eta-2}.\eea 

Note that if we were to use the approximation\cite{sur2019metallic,houghton2000multidimensional} where the dispersion in patch $\g$ is a function only of $\bfk\cdot\bfga$, we would not be able to reproduce the $\pi/4$ phase shift and the added factor of $1/2$ in the power law \eqref{uvbosoncorr}. 
Thus the $\psi$ correlator is sensitive to the smoothness of the Bose surface, and in order to obtain the correct correlation functions is essential to integrate over the whole Bose surface.

As a final example we can calculate the ``Kohn anomaly'' present at the fixed point, by examining how the correlation function of $\psi^\da \psi$ behaves at momenta with magnitude close to $2k_B$. In real space, we have 
\bea \lan (\psi^\da\psi) &(\bfx,\tau)  (\psi^\da\psi)(0)\ran \\  & \sim \frac{1}{N^2} \sum_{\g,\gp} \frac{ e^{ik_B x (\cos\g - \cos\gp)}}{[(\tau^2 + (x \cos \g)^2)(\tau^2 + (x\cos\gp)^2)]^{\eta/2}}. \eea 
We will be interested in the Fourier transform of this expression at zero frequency and at momentum with magnitude close to $2k_B$. Since we are using a UV cutoff at the length scale $\L\inv$, we will always have $k_Bx \gg1$ when Fourier transforming. Therefore again only the points of stationary phase ($\g,\gp = 0,\pi$) will contribute significantly to the angular integrals, allowing us to drop the $\g,\gp$ dependence in denominator. So then since $k\approx 2k_B$ also means $kx \gg 1$, the dominant part of the integral is 
\bea  \label{twobody_correlator} \lan \psi^\da\psi(\bfk,0) &  \psi^\da\psi(-\bfk,0)\ran \\ & \sim \int d\tau \, dx\, d\t\, d\g\, d\gp\, \frac{x e^{ix(-k \cos \t + k_B[\cos\g - \cos\gp])}}{(\tau^2 + x^2)^{\eta}}  \\ 
& \sim  \int dx\, \frac{\cos(kx-\pi/4) \cos^2(k_Bx - \pi/4)}{x^{2\eta-1/2}k^{1/2}} \\ & 
\sim \int dx \frac{\cos(x[k-2k_B] + \pi/4)}{x^{2\eta - 1/2 } k^{1/2}} \\ &
\sim {\rm Re}[(2k_B-k)^{2\eta - 3/2}] .\eea 
Note that if $\eta$ is such that the $\psi$ fields have the scaling dimensions of fermions $(\eta = 1)$, we get the same square root as in 2d Fermi liquids (recall that the interactions are irrelevant for $\eta>1$). 
In this case the singularity in \eqref{twobody_correlator} is one-sided and visible only at $k>2k_B$ (since the real part of $\sqrt{2k_B-k}$ then vanishes). For generic values of $\eta$ however the singularity is two-sided and visible for momentum transfer less than $k_B$.

\ss{Electromagnetic response \label{sec:background_fields}}

We now discuss the electromagnetic response of the BLL fixed point to determine if it is a superfluid, metal, or insulator. To do this we consider the response of the BLL phase to a background gauge field $A$ for the microscopic $U(1)$ symmetry, setting the Landau parameters to zero for simplicity. 

The background field enters the fixed-point action by coupling minimally to the $\phi_\g$ fields as\footnote{One way to double-check this expression is to re-write the Lagrangian in terms of the Fourier modes $\phi_l = \int \frac{d\g}\twp e^{il\g}\phi_\g$. Only $\phi_0$ is charged under the microscopic $U(1)$ symmetry, and so the theory can be gauged by minimally coupling $A$ to $\phi_0$. This gives the same answer as minimally coupling to the $\phi_\g$ fields directly; see appendix \ref{app:minimal_coupling} for details.} 
\bea \label{gauged_lagrangian} \mcl[A] = \frac{k_B}{\fpi  N \eta} \sum_\g \( v\inv (\p_\tau\phi_\g - A_\tau)^2  + v(\D_\g\phi_\g - \bfA\cdot\bfga)^2\).\eea 

We now integrate out the $\phi_\g$ fields to obtain the following effective Lagrangian for $A$:
\bea \mcl_{\rm eff}[A(\bfk,\o)] & = \frac{k_B}{\fpi \eta v} \int \frac{d\g}\twp \, A_\mu \Pi^{\mu\nu}_{T;\g} A_\nu,   \eea 
where $\Pi^{\mu\nu}_{T;\g}$ is the transverse projector in the spacetime plane $(\bfx\cdot\bfga,\tau )$. Explicitly,
\bea \label{emresponse} \mcl_{\rm eff}[A(\bfk,\o)] & = \frac{k_Bv}{\fpi \eta } \int \frac{d\g}\twp \frac{A_\tau^2 k_\g^2 + A_\g^2 \o^2 - 2A_\tau A_\g k_\g \o }{\o^2 + v^2k_\g^2}.\eea 
This expression is simplest in Coulomb gauge $\D \cdot \bfA = 0$, which we will adopt in what follows. Evaluating the integrals, we find 
\bea \label{integrated_em_response}
\mcl_{\rm eff}[A(\bfk,\o)] & = 
\frac{k_Bv}{\fpi \eta } \( \frac{A_\tau^2 \z^2 v^{-2}}{1+\z^2 + \sqrt{1+\z^2}} + \frac{\bfA^2 }{1+\sqrt{1+\z^2}} \),
\eea 
where we have defined $\z \equiv vk/\o$. 

Consider a scenario where $A_\tau=0$, with $\bfA$ tending to a constant. We can approach this in two limits, depending on whether we take $\o\rightarrow 0$ first followed by $k\ra 0$ or take the limit in the opposite order. The first limit corresponds to introducing a static transverse vector potential. A finite response in this limit implies Meissner screening and superfluidity. On the other hand, a finite response in the opposite order of limits only implies a finite Drude weight.\cite{scalapino1993insulator,resta2018drude}

If we first set $\o = 0$ and then take $k\ra0$, we send $\z \ra \infty$ in \eqref{integrated_em_response} and conclude that
\be \mcl_{\rm eff}[\bfA(\bfk \ra 0,\o=0)] = 0.\ee 
Therefore, like a Fermi liquid, the BLL has zero phase stiffness---thus there is no Meissner effect, and the BLL is not a superconductor.

If we now consider the opposite order of limits with $\z\ra0$, we see that 
\be \mcl_{\rm eff}[\bfA (\bfk = 0,\o \ra 0)] = \frac{k_Bv}{8\pi \eta} \bfA^2. \ee  
Therefore also like a Fermi liquid, the BLL has a finite Drude weight $D$, given by 
\be D = \frac{k_Bv}{4 \eta}.\ee 
Note that this is parametrically the same as the Fermi liquid result $D_{FL} = \pi n / m$ (in $e=1$ units), provided that we identify $m$ with $k_B / v$ and $n$ with $k_B^2/4\eta$. 
We conclude that the BLL is an example of a Bose metal. 

\section{BLLs in 3+1D \label{sec:threed}} 

In previous sections we have mostly focused on BLLs in 2+1D, but the generalization to 3+1D is straightforward. We consider the same type of Lagrangian as in \eqref{freeCircle}, with a dispersion possessing minima along a sphere of radius $k_B$. We then proceed by performing a patch decomposition of the Bose surface. We take each patch $\g$ to be a box of size $\L^3$ centered at $\bfga$, where now $\bfga$ lies on the unit $S^2$. The number of patches is accordingly 
\be N = \frac{4 \pi k_B^2}{\L^2}.\ee 

Following the same logic as in previous sections we arrive at the Lagrangian $\mcl_0 + \mcl_I$, with $\mcl_I$ containing symmetry-allowed interactions and with $\mcl_0$ given by 
\bea \label{threed_action} \mcl_0 & = \frac{k_B^2}{\fpi N \eta} \sum_{\g} \(v\inv (\p_\tau \phi_\g)^2  + v (\D_\g \phi_\g)^2\) \\ 
\mcl_f & = \frac{k_B^2}{\fpi N^2 \eta} \sum_{\g,\gp} \Big(v\inv f_\r^{\g,\gp} \p_\tau \phi_\g\p_\tau \phi_\gp   \\ & \qquad \qquad + v f_j^{\g,\gp}\D_\g \phi_\g \D_\gp \phi_\gp \Big).\eea  
The only differences with respect to the 2+1D action are the factors of $k_B^2$ up front (from dimensional analysis), and the fact that now the Landau parameters are functions of $z_{\bfga,\bfga'} \equiv \bfga\cdot\bfga' \in [-1,1]$. 

As in two dimensions, cosines in the dual variables $\t_\g$ are forbidden from appearing in $\mcl_I$. The most relevant term in $\mcl_I$ is again the BCS pairing interaction. Following the same logic as in section \ref{sec:2drg}, we write it as
\be \mcl_I \supset \frac{k_B^2 \L^2}{N^2} \sum_{\g,\gp} \ob g_{BCS}(z_{\g,\gp}) \cos(\vp_{\g,\gp}),\ee 
with $\ob g_{BCS}$ dimensionless,  and with $\vp_{\g,\gp}$ defined as before in \eqref{vphidef}. As in two dimensions the relevance of this term is found by comparing the dimension of $\cos(\vp_{\g,\g'})$ with $2$, so that as before the pairing interaction is irrelevant if $\eta > 1$.\footnote{ 
As in Fermi liquids, there are additional momentum-conserving two-body interactions present in three dimensions, known as non-forward scattering interactions.\cite{shankar1994renormalization} 
Using the RG framework of Ref. 
\onlinecite{lake2021fermi}, one can show that these interactions are always less relevant than the BCS pairing interaction, and as such can be ignored.}

The properties of the free fixed point \eqref{threed_action} are all rather similar to the 2+1D case. The vertex operators $e^{i\phi_\g}$ now have correlation functions 
\bea \label{threed_vertex_corrs} 
\lan e^{i\phi_\g(x)} e^{-i\phi_{\g'}(0)} \ran & \sim \d_{\g\g'}\d_\L(\bfx \cdot \bfga_{\perp,1}) \d_\L(\bfx \cdot \bfga_{\perp,2}) \\ & \qquad \times  \frac{1}{(1 +(\L v\tau)^2 + (\L \bfx \cdot \bfga)^2)^{\eta/2}},\eea 
where $\bfga_{\perp,1},\bfga_{\perp,2},\bfga$ constitute an orthonormal triad.
Similarly, the leading part of the equal-time UV boson correlation function at distances $x \gg k_B\inv$ is now 
\bea \chi(\bfx,0) & \sim  \int_{-1}^1 dz\,  \frac{e^{ik_B x z}}{(1+(\L x z)^2)^{\eta/2}} \\ & \sim \frac{\sin(k_Bx)}{x^{\eta+1}}.\eea 
The remaining aspects of the phenomenology can all be worked out in the same fashion as in section \ref{sec:2dpheno}.

\section{Electron transport in a BLL\label{sec:mnsi}}

In this section we discuss a situation wherein a metallic state of electrons coexists with a BLL. In such a setting, electron scattering off of the large density of low energy excitations of the BLL contributes to the resistivity, which we will show leads to an unusual temperature dependence of the form
\be \r \sim T^\eta,\ee 
where $\eta > 1$ is the exponent controlling correlation functions at the BLL fixed point. 

In particular, we will discuss a potential BLL arising in a metallic helimagnet in 3+1D, which may be realized for example in the B20 intermetallic compounds like MnSi and FeGe.\cite{pfleiderer1997magnetic,pfleiderer2004partial,Muhlbauer_2009,NFLMnSi,binz2006theory,nagaosaMnSi,FeGeNFL} We now briefly review the experimental situation in these systems, focusing on MnSi for concreteness. 

At ambient pressure, this system is a ferromagnetic metal, with a small Dzyaloshinskii–Moriya (DM) interaction favoring the development of long-wavelength spiral ordering in the magnetization.\cite{bak1980theory} The direction of the spiral ordering is determined by weak crystalline anisotropies, which pins the ordering along directions related by cubic symmetry.\cite{nakanishi1980origin,hopkinson2009origin}

As the pressure is increased, a first-order transition into a paramagnetic phase is observed.\cite{pfleiderer1997magnetic} This phase exhibits two remarkable properties. First, the spin degrees of freedom are seen to exhibit ``partial ordering'': the direction of the spiral ordering is no longer pinned, but the magnitude of the ordering wavevector remains well-defined, with neutron-scattering experiments seeing a nearly uniform intensity over a small sphere in momentum space.\cite{pfleiderer2004partial} Secondly, the resistivity is found to take on a non-Fermi liquid form, with $\r \sim T^{3/2}$ across the high-pressure phase.\cite{doiron2003fermi} In what follows we will see how both of these facts may be explained by modeling the spin fluctuations in the paramagnetic phase as a 3+1D BLL.  

To describe the spin fluctuations in the paramagnetic phase, we use a Landau-Ginzbarg Lagrangian for the magnetization vector $\bfM$, whose potential part quadratic in $\bfM$ contains the terms
\bea \label{dm_lagrangian} \mcl_M \supset  (\D \bfM)^2 + r \bfM^2 + 2 k_B \bfM \cdot (\D \times \bfM),\eea 
where the wavevector $k_B$ determines the strength of the DM interaction. To deal with the DM term, we follow Ref. \onlinecite{binz2006theory} and decompose the vector $\bfM$ into its constituent polarizations as 
\bea M^a(\bfx) & = \int  \dthq \, e^{i\bfq \cdot \bfx} \\ & \times \( \ep_1^a M_{l,\bfq} + \frac{\ep_2^a + i \ep_3^a}{\sqrt2} M_{+,\bfq} +  \frac{\ep_2^a - i \ep_3^a}{\sqrt2} M_{-,\bfq}\),\eea 
where $\bfep_1 = \bfq / q$, and where $\bfep_1,\bfep_2,\bfep_3$ constitute an orthonormal triad. Substituting this representation into \eqref{dm_lagrangian}, we see that the DM term becomes
\be 2k_B \bfM_\bfq^* \cdot (i\bfq \times \bfM_\bfq) = 2k_B q (|M_{+,\bfq}|^2 - |M_{-,\bfq}|^2).\ee 
The lowest energy mode is then $M_-$, which from now on we will write simply as $M$. Ignoring the higher-energy $M_l$ and $M_+$ modes, we then have 
\be \mcl_M \supset ((q-k_B)^2 - k_B^2 + r)|M_\bfq|^2,\ee 
so that the dispersion of $M$ has a degenerate minimum along a sphere of radius $k_B$. 

  Motivated by the fact that neutron scattering sees a nearly uniform intensity over a sphere in momentum space,\cite{pfleiderer2004partial} we make the assumption that the spin fluctuations can be captured by a 3+1D BLL formed from the negative polarization mode $M$, with the Bose surface being a sphere of radius $k_B$. 

We will now compute the consequences that this assumption has for the behavior of the itinerant electrons, which for simplicity we will take to form a Fermi gas with a spherical Fermi surface.
Including the coupling between the electrons and the spin fluctuations, the Lagrangian we are interested in is then
\bea \label{mnsi_lagrangian} \mcl  & = \mcl_M+ \mcl_c + \mcl_{cM}, \\
\mcl_c & = \ob c (\p_\tau - \ep_\bfk) c \\ 
\mcl_{cM} & = g \, \bar c_\a \s^a_{\a\b} c_\b \, M^a,  \eea 
where $\mcl_M$ is a BLL action for $M$ of the form written down in \eqref{threed_action}, $\ep_\bfk$ is the electron dispersion, and where the $M^a$ in $\mcl_{cM}$ implicity only contains the negatively polarized piece.
In what follows we will assume that the radius of the Bose surface is much smaller than that of the Fermi surface ($k_B / k_F \ll 1$), which is known to be the case in MnSi.\cite{pfleiderer2004partial}
 
The term $\mcl_{cM}$ will induce a finite scattering rate for the electrons. To determine this scattering rate, we will need to compute the contribution of the interaction term $\mcl_{cM}$ to the imaginary part of the electron self energy $\S$. The term which contributes to $\S$ at lowest order in $g$ is 
\be \S(\bfK,i\o_f) = g^2 T \sum_{\o_b} \int \dthq G_c(\bfK-\bfq,i(\o_f - \o_b)) \chi_M(\bfq,i\o_b),\ee 
where $\o_f$ and $\o_b$ are fermionic and bosonic Matsubara frequencies respectively, $G_c$ is the bare electron Greens function, and $\chi_M$ is the magnetic susceptibility of the BLL. 

The imaginary part of $\S$ is determined by employing a spectral representation for $G_c$ and $\chi_M$, with the spectral functions $A_c \equiv  - \pi\inv {\rm Im}[G_c]$ and $A_M \equiv - \pi\inv {\rm Im}[ \chi_M]$. Writing $G_c$ and $\chi_M$ in terms of $A_c$ and $A_M$, and resolving the Matsubara sum by integrating against the Bose distribution $n_B$, we have 
\bea \S(\bfK,i\o_f)  = - g^2 \int \dthq \, & d\O_1\, d\O_2 \,  (n_B(\O_2) - n_B(i\o_f - \O_1) ) \\ & \times  \frac{ A_c(\bfK- \bfq,\O_1) A_M(\bfq,\O_2) }{i\o_f - \O_1 - \O_2 + i\eta}. \eea 
Since $\o_f$ is a fermionic Matsubara frequency we may write $n_B(i\o_f - \O_1) = n_F(\O_1)-1$, with $n_F$ the Fermi function. Doing this and continuing to real frequencies, we then take the imaginary part and obtain 
\bea \S''(&\bfK,\o)   = \pi g^2 \int \dthq \,   d\O_1 \, d\O_2 \, \d(\o - \O_1 - \O_2) \,  \\ & \times  (1+ n_B(\O_2) - n_F(\O_1)) A_c(\bfK- \bfq,\O_1) A_M(\bfq,\O_2) .\eea 
Since the electrons are non-interacting in the absence of their coupling to the spin fluctuations, the electron spectral function is simply 
\be A_c(\bfk,\O) = \d(\O - \ep_\bfk),\ee 
so that 
\bea \S''(\bfK,\o) = \pi g^2 \int  & \dthq \,  (1+ n_B(\o - \ep_{\bfK-\bfq}) - n_F(\ep_{\bfK-\bfq})) \\ & \times A_M(\bfq,\o-\ep_{\bfK-\bfq}) .\eea 
The spectral function for the $M$ field is determined from the patch correlator \eqref{threed_vertex_corrs} after Fourier transforming and continuing to real frequencies as 
\be \label{bll_spectral_func} A_M(\bfq,\O) = A_0 \frac{\t(\O^2 - v^2 q_\prl^2)}{(\O^2 - v^2 q_\prl^2)^{\frac{2-\eta}{2}}}, \ee 
where $A_0$ is a constant and $q_\prl \equiv q - k_B$ as before. 

\begin{figure}
	\includegraphics[width=.4\textwidth]{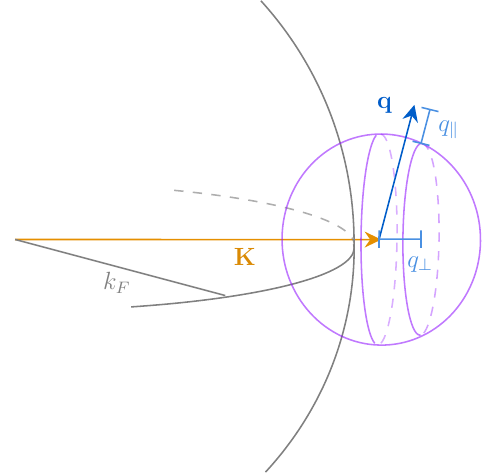} 
	\caption{\label{fig:two_surfaces} The geometry of the scattering processes contributing to $\S''$. A portion of the Fermi surface is drawn in gray, with the tip of the vector $\bfK$ lying just outside the Fermi surface. The purple sphere has radius $k_B$, and is centered on the tip of $\bfK$. The two vertical purple circles are separated by a distance of $k = |\bfK|-k_F$.  }
\end{figure}

We will first compute the $T=0$ scattering rate, working on-shell with $\o = \ep_\bfK$. We will take $\bfK$ to lie just outside the Fermi surface, with 
\be k \equiv |\bfK|-k_F \ll k_B \ll k_F.\ee 
In this case we have 
\bea \label{final_sigpp} \S''(\bfK,\ep_\bfK) = \pi g^2 \int \dthq \, &  \t(\ep_{\bfK-\bfq}) \t(\ep_\bfK- \ep_{\bfK-\bfq}) \\ & \times A_M(\bfq,\ep_\bfK - \ep_{\bfK - \bfq}).\eea

The region of momentum space contributing to this integral can be determined with the help of figure \ref{fig:two_surfaces}. The vector $\bfK$ is shown in orange, with its tip marking the origin of the coordinates for the $\bfq$ integral. The Fermi sphere is drawn in gray, and a sphere of radius $k_B$ is drawn in purple. Since $k_B \ll k_F$, we will approximate the Fermi surface as flat within a neighborhood of size $\sim k_B$ around $\bfK$. 
The first constraint comes from the $\t$ function in the $M$ spectral function \eqref{bll_spectral_func}, which  tells us that 
\be \ep_\bfK - \ep_{\bfK - \bfq} \approx v_F q_\perp > v q_\prl,\ee 
where $q_\perp \equiv \bfq \cdot \bfK / K$.
Second, the presence of the two $\t$ functions in \eqref{final_sigpp} restricts the integral over $\bfq$ to be such that $q_\perp < k$ and $0< q_\perp$. These two $\t$ functions restrict the range of $\bfq$ to the region in figure \ref{fig:two_surfaces} bounded by the two planes which intersect the purple sphere along the two vertical purple circles. 

We may now do the integral, which gives
\bea \S''(\bfK,\ep_\bfK) & = \pi g^2 A_0 \frac{\twp k_B}{(\twp)^3} \int_0^k dq_\perp  \\ & \qquad  \times  \int_0^{v_F q_\perp /v} dq_\prl  \, \frac1{((q_\perp v_F)^2 - (q_\prl v)^2)^{\frac{2-\eta}2}} \\ 
& = C  \frac{ g^2 k_B}{v_F v} \ep_k^\eta,\eea 
with $\ep_k = v_F k$ and with the constant 
\be C = \frac{A_0 \G(\eta/2) }{8 \sqrt\pi \eta \G((1+\eta)/2) }.\ee 
Since $\eta>1$ is needed for stability, the scattering rate vanishes faster than $\ep_k$ as $\bfK$ approaches the Fermi surface, and the quasiparticles remain well-defined. As such the electrons remain in a Fermi liquid state, albeit one with a faster scattering rate than in a conventional Landau Fermi liquid (provided that $\eta < 2$). 

To extract the transport lifetime of the quasiparticles from the above scattering rate, we need only multiply $\S''$ by $1-\cos\t$, where $\t$ is the typical scattering angle. In the present situation $\t \approx k_B / k_F \ll 1$, and so the transport scattering rate is 
\be  \G_{tr}(\ep,T=0)  = \frac{k_B^2}{2k_F^2} \S'' \propto \ep^\eta .\ee

Extending this result to finite $T$, the scattering rate is determined by scaling to be of the form 
\be \G_{tr}(\ep,T) = A T^\eta F\(\frac{\ep}{T}\),\ee 
where $A$ is a non-universal constant, and $F$ is a universal function. We thus obtain a contribution to the DC resistivity proportional to $T^\eta$, $\eta>1$. 

Of course in the present BLL + Fermi liquid theory the exponent $\eta$ is non-universal, and there is no a priori reason why it should take on the exact value of $3/2$ observed in experiments. However, a value of $\eta \approx 3/2$ is certainly possible, and as such the BLL + Fermi liquid model provides one possible explanation for the observed non-Fermi liquid behavior (with this explanation having the advantage of being particularly simple from an analytical standpoint). 

\section{Discussion \label{sec:disc}}

In this work we studied systems of translationally-invariant bosons (at both zero and nonzero densities) dubbed ``Bose-Luttinger liquids'' (BLLs). These phases of matter possess Bose surfaces and large emergent symmetry groups, and have regions of parameter space in which they are stable with respect to all symmetric perturbations. They lack quasiparticles and have continuously varying exponents, but also have phenomenology which is similar to Fermi liquids in some respects. There are many further questions that would be interesting to explore in future work. 

First, it would be desirable to have a better understanding of where BLLs are likely to show up in experiment. We have examined the example of MnSi in some detail, but it would be nice to explore other physical realizations further, such as pairing in FFLO superconductors and rotons in superfluid helium. 

In this paper we have only concerned ourselves with the phenomenology and stability of various BLL fixed points. One question to address is the ways in which the BLLs studied here can be connected to other phases of matter. As was already mentioned, one possibility is to study the transition driven by condensing vortices in the phase of the UV boson field. It would be interesting to understand how to perform this condensation in detail, as well as the nature of the resulting Mott insulating state one obtains in this way. 

One straightforward generalization of our work is to BLLs with generically-shaped Bose surfaces, beyond the simplest cases of the spherical Bose surfaces considered in the present work. As in Fermi liquids, the stability analysis of the IR theory will depend on the shape of the Bose surface, which will affect the types of symmetry-allowed perturbations to the fixed point one is allowed consider. 
It is also possible to consider fixed points where the anomalous dimension $\eta$ varies over the Bose surface. A scenario like this can occur if the momentum dependence of the microscopic interaction favors the average patch density $\r_\g$ to be a nontrivial function of $\g$, or if small rotation symmetry breaking terms are included in the dispersion of the UV bosons. 
Finally, it would be nice to have a more careful method of determining how the curvature of the Bose surface shows up in physical quantities and in RG flows, in a way which goes beyond the rather artificial patch construction employed here.

The BLLs constructed in this paper were approached by thinking of them as a large number of coupled Luttinger liquids. However, in principle one could imagine constructing IR theories out of other types of 1+1D CFTs, with one CFT living at each point on the Bose surface. At present it is not clear how exactly one would go about coupling the CFTs at each Bose surface point together, or whether there are any particular constraints on types of CFTs that can be chosen if the theory is to be regarded as arising from a UV lattice model of bosons. 

A final set of questions to address in future work relates to our treatment of the IR patch theory. First, it would be useful to have a more detailed understanding of when exactly our assumption about the uniform amplitude ordering of the $\psi_\g$ patch fields is justified. Secondly, it would be nice to find a way of dealing with the IR theory which doesn't rely on the patch decomposition used here --- within this framework a discussion of the emergent $LU(1)$ symmetry at the fixed point is rather awkward, as are issues relating to quantization and questions of duality between the phase and density fields. A more careful analysis of the field theories discussed here potentially would involve issues similar to those encountered in the analysis of the fractonic field theories studied in Refs. \cite{seiberg2020exoticI,seiberg2020exoticII,seiberg2020exoticIII,seiberg2020moreexotic}

\section*{Acknowledgements} EL thanks Zhen Bi, Dominic Else, Shu-Heng Shao, Ryan Thorngren, and Yizhi You for discussions. AV and TS would like to thank Benedikt Binz for a previous collaboration on closely related topics. EL is supported by the Fannie and John Hertz Foundation and the NDSEG fellowship. TS was supported by US Department of Energy grant DE- SC0008739, and partially through a Simons Investigator Award from the Simons Foundation. AV was supported through a Simons Investigator Award from the Simons Foundation. This work was also supported by the Simons Collaboration on Ultra-Quantum Matter, which is a grant from the Simons Foundation (651440, TS \&AV).
	
\appendix 
\section{Real bosons \label{app:real}}

In the main text we focused on theories of conserved complex bosons. 
One natural question to ask is whether the $U(1)$ charge conservation symmetry is in fact necessary for the realization of a stable BLL phase, or whether translation symmetry alone is sufficient.
This is an important question to address, as there are several scenarios in which we could imagine non-conserved bosons with the desired dispersion arising in experiment. 

One example is the superfluid phase of liquid He4, where the low-energy excitations are the sound mode and the roton. The latter has a dispersion possessing a minimum along a sphere in momentum space, and while while the roton gap $\De_R$ is finite in the superfluid phase, $\De_R$ is small and can be decreased by applying pressure. 
It is then perhaps not too outlandish to imagine a phase of He4 governed by a fixed point similar to the BLL described in the main text. 

\ss{1+1D}

As in the case of conserved bosons, it is easiest to warm up by looking at an example in 1+1D. 
We start by considering the Lagrangian 
\be \label{realmcl} \mcl = \frac12 \Psi\(-v\inv\p_\tau^2 + \frac{v}{4k_B^2}(-\p_x^2 - k_B)^2 + r\)\Psi + \frac g{24} \Psi^4.\ee
We will be interested in the regime where $r<0$. 

We start by breaking up $\Psi$ into left and right components 
\be \Psi_\lr(x) = \sqrt2 \int \dk e^{ikx} \Psi(\pm k_B + k),\ee 
where the integral is over an interval of length $2\L$. 
Due to the reality of $\Psi$ the left and right fields are not independent, and in fact constitute a single complex field 
\be \psi \equiv \Psi_L = \Psi_R^* .\ee  
In terms of $\psi$ the Lagrangian is then (after dropping irrelevant terms)
\be \mcl = \psi^*( -v\inv \p_\tau^2 - v \p_x^2 + r)\psi + \frac g4 |\psi|^4.\ee 
Therefore the IR theory is simply that of an XY model, with lattice-scale translations providing the $U(1)$ symmetry, which acts as $\psi \mt e^{ik_Ba}\psi$. 

The analysis then proceeds as in the case with complex bosons at zero density, except with half the number of fields due to the reality constraint. We work in terms of a phase field $\phi$ and its dual $\t$, with only $\phi$ transforming nontrivially under translation.  
Writing $\psi  = (r_0 + r)e^{ik_Bx + i\phi}$, the IR theory at energy scales below the mass of the $r$ field is 
\be \mcl = \frac{1}{4\pi \eta} \(v\inv (\p_\tau\phi)^2 + v(\p_x\phi)^2\) + g \cos(\t) + \cdots ,\ee 
where the $\cdots$ are less relevant symmetry-allowed interactions. 
When $\eta < 1/4$ the cosine is irrelevant and we obtain a single free boson, while for $\eta>1/4$ the IR theory is gapped and trivial. The only novelty about the gapless theory are $k_B$ oscillations in the correlation functions of the UV boson; otherwise the physics is simply that of the XY model. 

\ss{2+1D}

We now move on to 2+1D, and consider a UV Lagrangian of the form 
\bea \mcl & = \mcl_0 + \mcl_I \\
\mcl_0 &= \frac12(\p_\tau\Psi)^2 + \frac\b2 ((|\D| - k_B)\Psi)^2  + \frac r2 \Psi^2 \\ 
\mcl_I &= 
\frac\l6\Psi^3 + 
\frac g{24}\Psi^4,\eea 
with $r<0$. 
As usual, we define slowly fluctuating fields $\Psi_\g$ by breaking up $\Psi$ into $N$ momentum-space patches as  
\bea \Psi(\bfx,\tau) = \frac1{\sqrt N} \sum_\g \Psi_\g(\bfx,\tau) e^{ik_B \bfga \cdot \bfx},\eea 
with each patch of size $2\L\times 2\lap$.  
Each $\Psi_\g(\bfx,\tau)$ is a complex field, but the reality of $\Psi(\bfx,\tau)$ imposes the constraint 
\be \label{realconst} \Psi_\g(\bfx,\tau)^* = \Psi_{\g+\pi}(\bfx,\tau).\ee 
When written in terms of the $\Psi_\g$ fields the resulting IR Lagrangian is essentially the same as the the theory in \eqref{freeCircle}, but with two differences: the identification \eqref{realconst}, and the absence of a microscopic $U(1)$ symmetry. 

As usual, we now write $\Psi_\g \sim e^{i\Phi_\g}$. The reality constraint on the $\Phi_\g$ reads
\be \label{realconst_phase} \Phi_\g = - \Phi_{\g+\pi},\ee 
with the angle $\g$ now only running between $0$ and $\pi$. Translation symmetry acts on the $\Phi_\g$ as 
\be T_\mu \, : \, \cp_\g(\bfx) \mt \cp_\g(\bfx+\bfmu) +  k_B \bfmu \cdot \bfga.\ee

The analysis is then exactly the same as the zero-density limit of the previous theory where the UV bosons are complex, but with only half the number of IR fields. 
The IR Lagrangian can accordingly be written down as $\mcl_0 + \mcl_f + \mcl_I$, with 
\bea \label{iractionreal} \mcl_0 & = \frac{k_B }{\fpi N \eta} \sum_{\g=0}^\pi \(v\inv (\p_\tau \cp_\g)^2  + v (\D_\g \cp_\g)^2\) \\ 
\mcl_f & = \frac{k_B }{\fpi N^2 \eta} \sum_{\g,\gp=0}^\pi \Big(v\inv f_\r^{\g,\gp} \p_\tau \cp_\g\p_\tau \cp_\gp   \\ & \qquad \qquad+ v f_j^{\g,\gp}\D_\g \cp_\g \D_\gp \cp_\gp \Big),\eea 
and with $\mcl_I$ containing the symmetry-allowed interactions. As before, the coefficients appearing the above Lagrangian are all non-universal dimensionless numbers. The emergent symmetry of this fixed point is the subgroup of $LU(1)$ generated by odd angular momentum functions.

The most relevant translation-invariant cosine in the $\cp_\g$ variables is 
\be \label{gthree} \mcl_I \supset g_3\frac{k_B \L^2}{N} \sum_\g \cos(\cp_\g + \cp_{\g+\twp/3} + \cp_{\g-\twp/3}),\ee 
with $g_3$ a dimensionless coupling (there is no BCS-type term due to the constraint \eqref{realconst_phase}). Following the steps described insection \ref{sec:2drg}, we see that the RG eigenvalue of this perturbation is 
\be y_{g_3} = 2 - 3\eta /2,\ee 
and is therefore irrelevant if $\eta > 4/3$. 

As in the complex case, the absence of further relevant interactions depends on arguing that cosines in the dual fields $\ct_\g$ do not represent legitimate deformations to the fixed-point Lagrangian. The argument is essentially the same as in the complex case: the patch fields do not carry quantized charges (and indeed in the present setting there are no quantized charges at all, since the underlying degrees of freedom are real), and any putative patch vortex operator $e^{i\ct_\g}$ would create singular field configurations having infinite action in the presence of small perturbations containing derivatives along the Bose surface. 

The phenomenology of the fixed point \eqref{iractionreal} can be analyzed following the discussion ofsection \ref{sec:2dpheno}. Since there is no microscopic $U(1)$ symmetry there is no notion of compressibility, but the specific heat and real-space correlation functions all behave similarly to the complex case.

\section{Renormalization of $k_B$ \label{app:selfenergy}}

In the main text, the Bose momentum $k_B$ was defined simply via $\ep_0(k_B)=0$, with $\ep_0$ the non-interacting dispersion. 
In Fermi liquids renormalization of the Fermi momentum is forbidden at fixed density by Luttinger's theorem (at least in the presence of rotational symmetry), but of course here there is no analogous relation between $k_B$ and the boson density. The correct definition of $k_B$ in the presence of interactions is
\be \label{kbdef} \ep_0(k_B) - \Sigma(k_B,\o=0) = 0\ee 
with $\S(k,\o)$ the boson self-energy, meaning that interactions will generically renormalize $k_B$. 

More precisely, renormalization of $k_B$ will occur due to the reflection-odd terms appearing in the expansion for the dispersion close to the Bose surface. 
For example, consider the case of finite-density (non-relativistic) bosons in 1+1D, with the bosons at the right Bose point (say) having the dispersion 
\bea \label{1ddisp} \ep_0(k) & = \frac1{8mk_B^2}((k_B+k)^2-k_B^2)^2 \\ & =  \frac1{2m}\( k^2 + k^3/k_B + k^4/4k_B^2\),\eea 
where $k$ is the momentum relative to $k_B$. 
While the $k^3$ term is irrelevant, it will generate a term linear in $k$ under RG, which will then renormalize $k_B$. 
One may then worry that the renormalization is such that $k_B$ as defined in \eqref{kbdef} vanishes. If this happens then clearly the IR theory will not possess a Bose surface. 
In the following we will argue however that this renormalization can always be made small, so that there always exists a region of parameter space in which the Bose surface is stable. 

In section \ref{sec:finite_density}, we saw that in the case of 2+1D BLLs at finite density it was important to keep both the $\psi^* \p_\tau^2 \psi$ and $\psi^*\p_\tau \psi$ terms in the UV action. At large momenta the scaling about the UV fixed point will be governed by the relativistic $\psi^*\p_\tau^2 \psi$ term, while at lower momenta the scaling will be governed by the non-relativistic $\psi^*\p_\tau\psi$ term. The crossover between these regimes will happen at an intermediate scale $k_r$, which may or may not be larger than the scale at which the IR phase-only description sets in. 

In what follows we will consider scaling only either in the fully $z=1$ regime where the $\psi^*\p_\tau\psi$ term is absent (the particle-hole-symmetric limit of zero density), or in the fully $z=2$ regime where the $\psi^*\p_\tau^2 \psi$ term can be neglected for the purposes of computing the self energy in the UV scaling regime. This is done only for simplicity, and a more general analysis is possible.

\ss{$z=1$ scaling}  

Let us first discuss the renormalization of $k_B$ in the particle-hole symmetric limit of zero density, where the $\psi^*\p_\tau\psi$ term is absent. 
We will start by analyzing the 1+1D case, and will consider the Lagrangian 
\bea \label{relaction} \mcl  & =  \psi^*(-v\inv  \p_\tau^2 - v\p_x^2 -  i \z \p_x - i \b \p_x^3+m^2)\psi  + \frac g4 |\psi|^4,\eea
where the initial momentum cutoff is $\L \sim k_B$ and where $m^2<0$.
Here $\psi$ denotes the boson field expanded about one of the Bose points (for determining the renormalization of $k_B$ it is sufficient to focus on just a single Bose point), and we have dropped the $k^4$ term in \eqref{1ddisp} on the grounds that it is irrelevant and cannot generate a linear $k$ term by symmetry. We will start with $\z=0$ in the UV, but a nonzero $\z$ will be induced during the RG flow. 

If we write $\psi = (r_0+r) e^{i\phi}$, the IR theory we are aiming for is one written only in terms of the field $\phi$. The IR regime is then determined by the scale at which we can neglect fluctuations in the radial mode $r$ and focus only on the field $\phi$. This happens approximately when the cutoff reaches the mass $m_r$ of the $r$ field, which occurs after an RG time of 
\be t_c \approx \ln(k_B / m_r).\ee 
The Bose surface will thus be stable if by an RG time of $t_c$ the renormalization of $k_B$ satisfies $\d k_B \ll m_r$: if this is the case we will be left with a theory with a dispersion with minima at $\pm (k_B + \d k_B)$, and the $k^3$ term will cease to renormalize $k_B$ since $\phi \p^3 \phi$ is a total derivative. By plugging in the polar representation of $\psi$ into the above action and solving for the $r_0$ which minimizes the potential, we see that 
\be m_r = \sqrt{g\rho_S} = k_B\sqrt{\bar g(0) \r_S},\ee 
where $\r_S = r_0^2 = -2m^2/g$ is the boson amplitude evaluated at the classical minimum of the potential in \eqref{relaction}, and where $\bar g(0) = gk_B^{-2}$ is the bare interaction strength made dimensionless with the UV cutoff $\L \sim k_B$. 

To get an idea for how the $\b k^3$ term contributes to the self-energy, we break up $\psi$ as $\psi = \sqrt{\r_S} + \chi$, with $\chi$ a field parametrizing the fluctuations about the classical minimum. The action is now (in $v=1$ units and treating $\chi$ and its conjugate $\bar\chi$ as two separate fields)
\begin{widetext}
\be\label{chiaction} S = \int \dkdo \, \(\frac12\bpm \bar\chi \, \chi\epm_{-k,-\o} \mcg\inv(k,\o) \bpm \bar\chi \\ \chi\epm_{k,\o} + \frac{g\sqrt{\r_S}}2 ((\bar\chi)^2 \chi + \bar\chi \chi^2) + \frac{g}{4} |\chi|^4\),\ee  
with the propagator 
\bea \mcg(k,\o) &= \frac{1}{(\o^2+k^2)(\o^2+k^2-2m^2) - \b^2 k^6} \bpm m^2 & \o^2+k^2-m^2 + \b k^3 \\ \o^2+k^2-m^2 - \b k^3 & m^2 \epm, \eea 
which when rotated to real frequencies has poles at the correct locations $\o = k, \sqrt{k^2+m_r^2}$.

\begin{figure}
	\centering 
	\includegraphics{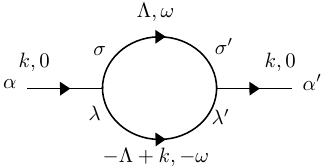}
	\caption{\label{oneloop} The 1-loop diagrams contributing to $\S_{\a\a'}(\bfk,0)$ for $\bfk>0$.}
\end{figure}

We will compute the flow of the self energy $\S(k,0)$ by integrating out modes of all frequencies, and with momentum lying in an interval of width $d\L$ about the points $\pm \L$. 
The flow of the zero-frequency self-energy $\S_{\a\a'}(k,0)$ (where $\a,\a' \in \{\chi,\bar\chi\}$) is then given to 1-loop order by 
\bea \frac{d\S_{\a\a'}(k,0)}{d\L} & = 2g^2 \r_S \int \frac{d\o}{\twp} \sum_{\s\s'\l\l' = \bar\chi,\chi} V^\a_{\s\l} V^{\a'}_{\s'\l'} \( \ct(k) \mcg_{\s\s'}(\L,\o)\mcg_{\l\l'}(-\L + |k|,-\o) + \ct(-k) \mcg_{\s\s'}(-\L,-\o)\mcg_{\l\l'}(\L-|k|,\o)\), \eea  
where we have defined the factors $V^\a_{\b\g}$, which equal zero if $\a=\b=\g$ and are equal to $1$ otherwise. The diagram corresponding to the first term proportional to $\ct(k)$ is shown in Figure \ref{oneloop}.

\begin{figure}
	\includegraphics{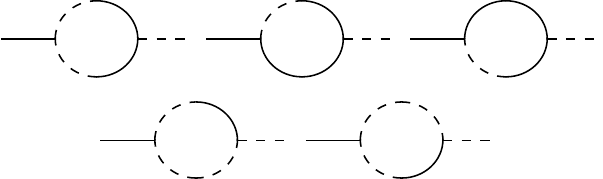}
	\caption{\label{odddiagrams} The diagrams which contribute to the part of $\Sigma_{\chi\bar\chi}(k,0)$ odd in $k$. Solid lines represent $\mcg_{\chi\chi}$, dashed lines represent $\mcg_{\bar\c\bar\c}$, and mixed dashed-solid lines represent $\mcg_{\bar\c\c},\mcg_{\c\bar\c}$. } 
\end{figure}
Since $\mcg_{\s\s'}(q,\nu) = \mcg_{\s'\s}(-q,-\nu)$, we can re-write the above as 
\bea \frac{d\S_{\a\a'}(k,0)}{d\L} & = 2g\r_S\int \frac{d\o}{\twp} \sum_{\s\s'\l\l' = \bar\chi,\chi} \mcg_{\s\s'}(\L,\o)\mcg_{\l\l'}(-\L+|k|,-\o) \( \ct(k) V^\a_{\s\l} V^{\a'}_{\s'\l'} + \ct(-k) V^{\a'}_{\s\l} V^{\a}_{\s'\l'}\).\eea 
\end{widetext} 
We therefore see that for diagrams in which the two $VV$ factors above are equal, the change in self energy is in fact only a function of $|k|$, and hence cannot contribute to a renormalization of $k_B$. As such the $\S_{\chi\chi}$ and $\S_{\bar\chi\bar\chi}$ components of the self energy do not pick up any terms odd in $k$ under the momentum shell integration, and therefore we may focus on the flow of $\S_{\chi\bar\chi}$.

By the same reasoning only a subset of the 1-loop diagrams appearing in $\S_{\chi\bar\chi}$ contribute to the $k$-odd part of $d\S_{\chi\bar\chi}/d\L$; these diagrams are those for which $V^\chi_{\s\l} V^{\bar\chi}_{\s'\l'} \neq V^{\bar\chi}_{\s\l} V^\chi_{\s'\l'}$, and are shown in Figure \ref{odddiagrams}. 

The full evaluation of these diagrams is rather complicated, but we will only concern ourselves with the part linear in $k/\L$, first order in $\L\b$, and lowest order in $m_r/\L$. We then find that the flow of the $\sgn(k)$-dependent part of $\S_{\bar\chi\chi}$ is given by 
\be \frac{d\S_{\bar\c\c}}{dt} \supset \sfd    \r_S \bar \b\bar g^2 k \L ,\ee 
where $\bar g = g^2 / \L^2$ and $\bar\b = \b \L$ are dimensionless couplings and $\sfd$ is a positive constant. 
If we then let $\bar\z = \z /\L$ denote the dimensionless coupling associated to the linear $k$ term, we then get the leading order flow 
\be \frac{d\bar\z}{dt} \approx\bar\z + \sfd \r_S \bar\b(t) \bar g^2(t).\ee 
In the first stages of the flow we have $\bar g(t) \approx \bar g(0)e^{2t}$ and $\bar \b(t) \approx \bar\b(0)e^{-t}$. Starting from $\bar\z(0) = 0$ and flowing up to $t_c$, we then get 
\bea \bar\z(t_c) & \approx \frac{\sfd\r_Sk_B}{2m_r}  \bar\b(0)\bar g(0)^2 ((k_B/m_r)^2-1). \eea 
In order for this treatment of the flow of $\bar\z$ to remain valid, $\bar g(t)$ must be small for all $t$ along the flow. This means that we must have 
\be \( \frac{k_B}{m_r}\)^2 \bar g(0) \lesssim 1 \implies \r_S \gtrsim 1.\ee 
As long as this condition is satisfied,  $\bar\z(t_c)$ can be made arbitrarily small by taking the dimensionless parameter $\bar g(0) \ra 0$, and thus there exists a regime of parameter space in which we expect the Bose surface to be stable. 

While the above analysis was done in 1+1D, the conclusion is unchanged in 2+1D. In 2+1D, we are interested in the $k$-dependence of the patch self-energy $\S_\g(k,0)$. Since the self-energy for each patch field is of order $1/N = \L / \pi k_B$ (for the same reason that the Landau parameters only affect the patch propagators at order $1/N$), $k_B$ will cease to renormalize once we reach cutoffs $\L$ such that $N\gg 1$. The renormalization of $k_B$ during the early parts of the flow where $N$ is of order 1 can be argued to be arbitrarily small using a similar argument as in 1+1D, and we conclude that interactions do not necessarily destabilize the Bose surface. 

\ss{$z=2$ scaling}

We now address the limit where the $\psi^*\p_\tau^2\psi$ term in the action can be neglected for the purposes of computing the flow of the linear $k$ term. In 1+1D, the Lagrangian we are interested in is 
\be \label{z2_oned_action}\mcl =  \psi^*(i\p_t + \mu + \frac1{2m} \z \, i\p_x  + \frac1{2m}\p_x^2 - i \b \p_x^3)\psi - \frac g4 |\psi|^4,\ee 
where $\z$ has units of momentum and is taken to have a vanishing bare value, and where $\mu,g>0$ set the average density. 

 We again write $\psi$ as $\psi = \sqrt{\bar\r} + \chi$, with $\chi$ a complex field capturing the fluctuations about the background density. The action is then the same as \eqref{chiaction}, except that the propagator is now (in real time)
\bea \, & \mcg(k,\o)  = \frac i{(\o + \b k^3)^2 - \xi_k^2 + i0} \\ & \qquad  \times  \bpm -\mu &\o + \b k^3 + k^2 /2m + \mu \\ -(\o + \b k^3) + k^2/2m + \mu & -\mu \epm,\eea
with 
\be \xi_k^2 \equiv \frac{ k^2}{2m}\(\frac{ k^2}{2m} + \mu\).\ee 
When $\b=0$ we check that $\mcg$ has poles at $\pm \xi_k$, correctly giving the Bogoliubov dispersion. 

The crossover between the high-momentum non-relativistic $\o \approx k^2/2m$ and the low-momentum relativistic $\o \approx \sqrt{\mu/2m} k$ occurs at the momentum $k_c = \sqrt{2m\mu}$, which in terms of the average density $\bar \r$ is 
\be k_c = \sqrt{mg\bar \r}.\ee 
At this scale, the behavior crosses over from massive particles in the UV to coherent waves in the IR, where the hydrodynamic phase-only representation sets in. 

In order to reach the hydrodynamic regime in the IR, we then need to flow for an RG time of $t_c \approx \ln(k_B / k_c)$. 
The 1-loop diagrams contributing to the part of the self-energy sensitive to the sign of $k$ are the same as in the previous section. 
To first order in $m\L\b$ and $k / \L$, we find 
\be \frac{d\S_{\bar\c\c}}{dt} \supset  \sfc \bar\b \bar g^2 \frac{k \bar\r}{m},\ee 
where the dimensionless parameters are now defined as 
\be \bar g = \frac{m g}{\L},\qquad \bar\b = m\L \b,\ee 
and where $\sfc$ is a positive constant. 
$\bar\beta$ is irrelevant with RG eigenvalue $-1$ under the UV scaling, while $\bar g$ is relevant with RG eigenvalue $+1$. In terms of the parameter $\z$ in \eqref{z2_oned_action}, we see that $d\z / dt \approx \z + \sfc \bar \beta \bar g^2 \bar \r$. At an RG time $t$, $\z$ is then approximately 
\be \z(t) \approx \sfc \bar\beta(0)\bar g(0)^2 \bar\r e^t t.\ee 
The hydrodynamic scaling regime is reached after a time $t_c\approx \ln(k_B/k_c)$, at which point the effective dispersion is approximately $k^2/2m - k \z(t_c)/2m$, which gives a renormalization of $k_B$ by an amount $\d k_B \approx \z(t_c)/2$. In order for the Bose surface to be stable then, we require that $\z(t_c) / k_c \ll 1$. Now using $k_c \approx \sqrt{\bar g(0)k_B \bar \r}$, we have 
\bea \frac{\z(t_c)}{k_c}  & \approx \sfc \bar\b(0)\bar g(0)^2 \ln(k_B/k_c) \frac{k_B\bar\r}{k_c^2} \approx \frac\sfc2 \bar\b(0) \bar g(0) \ln \(\frac{k_B}{\bar\r \bar g(0)}\).\eea 
Thus $\z(t_c)/k_c$ can be made arbitrarily small if $\bar g(0)$ is made small.

\section{A more careful justification of minimal coupling \label{app:minimal_coupling}}

In this appendix we provide a justification for the claim that the 2+1D BLL considered in the main text may be coupled to a gauge field $A$ for the microscopic $U(1)$ symmetry by minimally coupling $A$ to the $\phi_\g$ fields on each patch, as was written down in \eqref{gauged_lagrangian}. 

Our motivation for critically examining the minimal coupling of \eqref{gauged_lagrangian} can be understood by thinking about what happens in the context of Fermi liquids. 
In the bosonized description of Fermi liquids, the chirality of the patch fields means that it is incorrect to minimally couple $A$ to the phase fields on each patch. Instead, the correct thing to do\cite{mross2011decohering} is to re-write the Lagrangian in terms of the Fourier modes 
\be \phi_l = \int \frac{d\g}\twp  e^{il\g} \phi_\g,\ee 
and then to minimally couple $A$ to $\phi_0$. Indeed, as was discussed in the main text, the $\phi_\g$ are not independent $\twp$ periodic variables---the only periodic variable is $\phi_0$, and so only $\phi_0$ should couple to $A$. 

Unlike the Fermi liquid the BLL is not chiral, and this means the naive minimal coupling in \eqref{gauged_lagrangian} is indeed correct. It is however worthwhile to demonstrate this fact explicitly. 

Working in the zero density limit, and setting the Landau parameters to zero for simplicity, we can write the IR Lagrangian as 
\begin{widetext}
\bea \mcl & = \sfk \int_\g \( (\p_\tau \phi_\g)^2 + (\D_\g \phi_\g)^2 \) \\ 
& =\sfk \sum_{n,m} \( \p_\tau \phi_n \p_\tau \phi_m \d_{n,-m} + \int_\g (\cos(\g) \p_x + \sin(\g) \p_y)\phi_n (\cos(\g) \p_x + \sin(\g)\p_y )\phi_m e^{i\g(n+m)}\)  \\ 
& = - \sfk  \sum_n \phi_n \Big[ \p_\tau^2 \phi_{-n}  + \frac14\Big( \p_x^2( 2\phi_{-n} + \phi_{-n-2} + \phi_{-n+2}) + \p_y^2(2\phi_{-n} - \phi_{-n-2} - \phi_{-n+2}) + 2i\p_x\p_y(\phi_{-n+2} -\phi_{-n-2}) \Big) \Big] \eea 
where for convenience we have defined $\sfk = k_B / \fpi \eta$ and $\int_\g = \int \frac{d\g}\twp$. The part of $\mcl$ containing $\phi_0$ is 
\bea \mcl[\phi_0] & = -\sfk \[ \phi_0\(\p_\tau^2 + \frac12\D^2\)\phi_0 + \frac12 \phi_0\( (\p_x + i\p_y)^2 \phi_2 + (\p_x - i \p_y)^2 \phi_{-2}\)\]\eea 
with the couplings to $\phi_{\pm2}$ taking the correct form required by rotational symmetry. 

If we now couple $\phi_0$ minimally to $A$, the terms dependent on $A$ are 
\bea \mcl[A] & = \sfk \[ -2\p_\tau \phi_0 A_\tau + A_\tau^2 - \bfA\cdot \D \phi_0 + \frac12 \bfA^2 - \frac12 (A_x+iA_y)  (\p_x+i\p_y)\phi_2 - \frac12 (A_x-iA_y)(\p_x-i\p_y) \phi_{-2} \].\eea

If we now rewrite this in terms of the patch fields, we find 
\bea \mcl[A] & = \sfk \int_\g \[ A_\tau^2  + \frac12 \bfA^2 + \( -2 A_\tau \p_\tau - \bfA\cdot \D - \cos(2\g)(A_x \p_x - A_y\p_y) - \sin(2\g)(A_x\p_y + A_y\p_x)\)\phi_\g\] \\ 
& = \sfk \int_\g \[ A_\tau^2 + A_\g^2 + \(  -2A_\tau\p_\tau- 2 \cos^2(\g) A_x\p_x - 2 \sin^2(\g) A_y \p_y -2 \cos(\g)\sin(\g) (A_x\p_y + A_y\p_x) \)\phi_\g\] \\ 
& = \sfk \int_\g \[ A_\tau^2+ A_\g^2  - 2 \( A_\tau\p_\tau + A_\g \D_\g\) \phi_\g\],  \eea 
which has exactly the same $A$ dependence as the naive minimal coupling in \eqref{gauged_lagrangian}. 

\,\\ 

\end{widetext}

\bibliography{bose_liquids_revisited}

\begin{thebibliography}{45}
\expandafter\ifx\csname natexlab\endcsname\relax\def\natexlab#1{#1}\fi
\expandafter\ifx\csname bibnamefont\endcsname\relax
  \def\bibnamefont#1{#1}\fi
\expandafter\ifx\csname bibfnamefont\endcsname\relax
  \def\bibfnamefont#1{#1}\fi
\expandafter\ifx\csname citenamefont\endcsname\relax
  \def\citenamefont#1{#1}\fi
\expandafter\ifx\csname url\endcsname\relax
  \def\url#1{\texttt{#1}}\fi
\expandafter\ifx\csname urlprefix\endcsname\relax\def\urlprefix{URL }\fi
\providecommand{\bibinfo}[2]{#2}
\providecommand{\eprint}[2][]{\url{#2}}

\bibitem[{\citenamefont{Das and Doniach}(1999)}]{das1999existence}
\bibinfo{author}{\bibfnamefont{D.}~\bibnamefont{Das}} \bibnamefont{and}
  \bibinfo{author}{\bibfnamefont{S.}~\bibnamefont{Doniach}},
  \bibinfo{journal}{Physical Review B} \textbf{\bibinfo{volume}{60}},
  \bibinfo{pages}{1261} (\bibinfo{year}{1999}).

\bibitem[{\citenamefont{Phillips and Dalidovich}(2003)}]{phillips2003elusive}
\bibinfo{author}{\bibfnamefont{P.}~\bibnamefont{Phillips}} \bibnamefont{and}
  \bibinfo{author}{\bibfnamefont{D.}~\bibnamefont{Dalidovich}},
  \bibinfo{journal}{Science} \textbf{\bibinfo{volume}{302}},
  \bibinfo{pages}{243} (\bibinfo{year}{2003}).

\bibitem[{\citenamefont{Dalidovich and
  Phillips}(2001)}]{dalidovich2001interaction}
\bibinfo{author}{\bibfnamefont{D.}~\bibnamefont{Dalidovich}} \bibnamefont{and}
  \bibinfo{author}{\bibfnamefont{P.}~\bibnamefont{Phillips}},
  \bibinfo{journal}{Physical Review B} \textbf{\bibinfo{volume}{64}},
  \bibinfo{pages}{052507} (\bibinfo{year}{2001}).

\bibitem[{\citenamefont{Dalidovich and Phillips}(2002)}]{dalidovich2002phase}
\bibinfo{author}{\bibfnamefont{D.}~\bibnamefont{Dalidovich}} \bibnamefont{and}
  \bibinfo{author}{\bibfnamefont{P.}~\bibnamefont{Phillips}},
  \bibinfo{journal}{Physical review letters} \textbf{\bibinfo{volume}{89}},
  \bibinfo{pages}{027001} (\bibinfo{year}{2002}).

\bibitem[{\citenamefont{Paramekanti et~al.}(2002)\citenamefont{Paramekanti,
  Balents, and Fisher}}]{paramekanti2002ring}
\bibinfo{author}{\bibfnamefont{A.}~\bibnamefont{Paramekanti}},
  \bibinfo{author}{\bibfnamefont{L.}~\bibnamefont{Balents}}, \bibnamefont{and}
  \bibinfo{author}{\bibfnamefont{M.~P.} \bibnamefont{Fisher}},
  \bibinfo{journal}{Physical Review B} \textbf{\bibinfo{volume}{66}},
  \bibinfo{pages}{054526} (\bibinfo{year}{2002}).

\bibitem[{\citenamefont{Ma and Pretko}(2018)}]{ma2018higher}
\bibinfo{author}{\bibfnamefont{H.}~\bibnamefont{Ma}} \bibnamefont{and}
  \bibinfo{author}{\bibfnamefont{M.}~\bibnamefont{Pretko}},
  \bibinfo{journal}{Physical Review B} \textbf{\bibinfo{volume}{98}},
  \bibinfo{pages}{125105} (\bibinfo{year}{2018}).

\bibitem[{\citenamefont{Seiberg and
  Shao}(2020{\natexlab{a}})}]{seiberg2020exoticI}
\bibinfo{author}{\bibfnamefont{N.}~\bibnamefont{Seiberg}} \bibnamefont{and}
  \bibinfo{author}{\bibfnamefont{S.-H.} \bibnamefont{Shao}},
  \bibinfo{journal}{arXiv preprint arXiv:2003.10466}
  (\bibinfo{year}{2020}{\natexlab{a}}).

\bibitem[{\citenamefont{Seiberg and
  Shao}(2020{\natexlab{b}})}]{seiberg2020exoticII}
\bibinfo{author}{\bibfnamefont{N.}~\bibnamefont{Seiberg}} \bibnamefont{and}
  \bibinfo{author}{\bibfnamefont{S.-H.} \bibnamefont{Shao}},
  \bibinfo{journal}{arXiv preprint arXiv:2004.00015}
  (\bibinfo{year}{2020}{\natexlab{b}}).

\bibitem[{\citenamefont{Tay et~al.}(2011)\citenamefont{Tay, Motrunich
  et~al.}}]{tay2011possible}
\bibinfo{author}{\bibfnamefont{T.}~\bibnamefont{Tay}},
  \bibinfo{author}{\bibfnamefont{O.~I.} \bibnamefont{Motrunich}},
  \bibnamefont{et~al.}, \bibinfo{journal}{Physical Review B}
  \textbf{\bibinfo{volume}{83}}, \bibinfo{pages}{205107}
  (\bibinfo{year}{2011}).

\bibitem[{\citenamefont{Xu and Fisher}(2007)}]{xu2007bond}
\bibinfo{author}{\bibfnamefont{C.}~\bibnamefont{Xu}} \bibnamefont{and}
  \bibinfo{author}{\bibfnamefont{M.~P.} \bibnamefont{Fisher}},
  \bibinfo{journal}{Physical Review B} \textbf{\bibinfo{volume}{75}},
  \bibinfo{pages}{104428} (\bibinfo{year}{2007}).

\bibitem[{\citenamefont{You et~al.}(2020)\citenamefont{You, Bi, and
  Pretko}}]{you2020emergent}
\bibinfo{author}{\bibfnamefont{Y.}~\bibnamefont{You}},
  \bibinfo{author}{\bibfnamefont{Z.}~\bibnamefont{Bi}}, \bibnamefont{and}
  \bibinfo{author}{\bibfnamefont{M.}~\bibnamefont{Pretko}},
  \bibinfo{journal}{Physical Review Research} \textbf{\bibinfo{volume}{2}},
  \bibinfo{pages}{013162} (\bibinfo{year}{2020}).

\bibitem[{\citenamefont{Xu and Moore}(2005)}]{xu2005reduction}
\bibinfo{author}{\bibfnamefont{C.}~\bibnamefont{Xu}} \bibnamefont{and}
  \bibinfo{author}{\bibfnamefont{J.}~\bibnamefont{Moore}},
  \bibinfo{journal}{Nuclear Physics B} \textbf{\bibinfo{volume}{716}},
  \bibinfo{pages}{487} (\bibinfo{year}{2005}).

\bibitem[{\citenamefont{Anderson}(1990)}]{anderson1990luttinger}
\bibinfo{author}{\bibfnamefont{P.~W.} \bibnamefont{Anderson}},
  \bibinfo{journal}{Physical review letters} \textbf{\bibinfo{volume}{64}},
  \bibinfo{pages}{1839} (\bibinfo{year}{1990}).

\bibitem[{\citenamefont{Doiron-Leyraud
  et~al.}(2003)\citenamefont{Doiron-Leyraud, Walker, Taillefer, Steiner,
  Julian, and Lonzarich}}]{doiron2003fermi}
\bibinfo{author}{\bibfnamefont{N.}~\bibnamefont{Doiron-Leyraud}},
  \bibinfo{author}{\bibfnamefont{I.}~\bibnamefont{Walker}},
  \bibinfo{author}{\bibfnamefont{L.}~\bibnamefont{Taillefer}},
  \bibinfo{author}{\bibfnamefont{M.}~\bibnamefont{Steiner}},
  \bibinfo{author}{\bibfnamefont{S.}~\bibnamefont{Julian}}, \bibnamefont{and}
  \bibinfo{author}{\bibfnamefont{G.}~\bibnamefont{Lonzarich}},
  \bibinfo{journal}{Nature} \textbf{\bibinfo{volume}{425}},
  \bibinfo{pages}{595} (\bibinfo{year}{2003}).

\bibitem[{\citenamefont{Sur and Yang}(2019)}]{sur2019metallic}
\bibinfo{author}{\bibfnamefont{S.}~\bibnamefont{Sur}} \bibnamefont{and}
  \bibinfo{author}{\bibfnamefont{K.}~\bibnamefont{Yang}},
  \bibinfo{journal}{Physical Review B} \textbf{\bibinfo{volume}{100}},
  \bibinfo{pages}{024519} (\bibinfo{year}{2019}).

\bibitem[{\citenamefont{Po and Zhou}(2015)}]{po2015two}
\bibinfo{author}{\bibfnamefont{H.~C.} \bibnamefont{Po}} \bibnamefont{and}
  \bibinfo{author}{\bibfnamefont{Q.}~\bibnamefont{Zhou}},
  \bibinfo{journal}{Nature communications} \textbf{\bibinfo{volume}{6}},
  \bibinfo{pages}{1} (\bibinfo{year}{2015}).

\bibitem[{\citenamefont{Neto and Fradkin}(1994)}]{neto1994bosonization}
\bibinfo{author}{\bibfnamefont{A.~C.} \bibnamefont{Neto}} \bibnamefont{and}
  \bibinfo{author}{\bibfnamefont{E.}~\bibnamefont{Fradkin}},
  \bibinfo{journal}{Physical Review B} \textbf{\bibinfo{volume}{49}},
  \bibinfo{pages}{10877} (\bibinfo{year}{1994}).

\bibitem[{\citenamefont{Houghton et~al.}(2000)\citenamefont{Houghton, Kwon, and
  Marston}}]{houghton2000multidimensional}
\bibinfo{author}{\bibfnamefont{A.}~\bibnamefont{Houghton}},
  \bibinfo{author}{\bibfnamefont{H.-J.} \bibnamefont{Kwon}}, \bibnamefont{and}
  \bibinfo{author}{\bibfnamefont{J.}~\bibnamefont{Marston}},
  \bibinfo{journal}{Advances in Physics} \textbf{\bibinfo{volume}{49}},
  \bibinfo{pages}{141} (\bibinfo{year}{2000}).

\bibitem[{\citenamefont{Froehlich and
  Goetschmann}(1997)}]{froehlich1997bosonization}
\bibinfo{author}{\bibfnamefont{J.}~\bibnamefont{Froehlich}} \bibnamefont{and}
  \bibinfo{author}{\bibfnamefont{R.}~\bibnamefont{Goetschmann}},
  \bibinfo{journal}{Physical Review B} \textbf{\bibinfo{volume}{55}},
  \bibinfo{pages}{6788} (\bibinfo{year}{1997}).

\bibitem[{\citenamefont{Shankar}(1994)}]{shankar1994renormalization}
\bibinfo{author}{\bibfnamefont{R.}~\bibnamefont{Shankar}},
  \bibinfo{journal}{Reviews of Modern Physics} \textbf{\bibinfo{volume}{66}},
  \bibinfo{pages}{129} (\bibinfo{year}{1994}).

\bibitem[{\citenamefont{Brazovskiǐ}(1975)}]{brazovskii1975phase}
\bibinfo{author}{\bibfnamefont{S.}~\bibnamefont{Brazovskiǐ}},
  \bibinfo{journal}{JETP} \textbf{\bibinfo{volume}{41}}, \bibinfo{pages}{85}
  (\bibinfo{year}{1975}).

\bibitem[{\citenamefont{Pisarski et~al.}(2020)\citenamefont{Pisarski, Tsvelik,
  and Valgushev}}]{pisarski2020transverse}
\bibinfo{author}{\bibfnamefont{R.~D.} \bibnamefont{Pisarski}},
  \bibinfo{author}{\bibfnamefont{A.~M.} \bibnamefont{Tsvelik}},
  \bibnamefont{and}
  \bibinfo{author}{\bibfnamefont{S.}~\bibnamefont{Valgushev}},
  \bibinfo{journal}{Physical Review D} \textbf{\bibinfo{volume}{102}},
  \bibinfo{pages}{016015} (\bibinfo{year}{2020}).

\bibitem[{\citenamefont{Binz et~al.}(2006)\citenamefont{Binz, Vishwanath, and
  Aji}}]{binz2006theory}
\bibinfo{author}{\bibfnamefont{B.}~\bibnamefont{Binz}},
  \bibinfo{author}{\bibfnamefont{A.}~\bibnamefont{Vishwanath}},
  \bibnamefont{and} \bibinfo{author}{\bibfnamefont{V.}~\bibnamefont{Aji}},
  \bibinfo{journal}{Physical review letters} \textbf{\bibinfo{volume}{96}},
  \bibinfo{pages}{207202} (\bibinfo{year}{2006}).

\bibitem[{\citenamefont{Haldane}(2005)}]{haldane2005luttinger}
\bibinfo{author}{\bibfnamefont{F.}~\bibnamefont{Haldane}},
  \bibinfo{journal}{arXiv preprint cond-mat/0505529}  (\bibinfo{year}{2005}).

\bibitem[{\citenamefont{Else et~al.}(2020)\citenamefont{Else, Thorngren, and
  Senthil}}]{else2020non}
\bibinfo{author}{\bibfnamefont{D.~V.} \bibnamefont{Else}},
  \bibinfo{author}{\bibfnamefont{R.}~\bibnamefont{Thorngren}},
  \bibnamefont{and} \bibinfo{author}{\bibfnamefont{T.}~\bibnamefont{Senthil}},
  \bibinfo{journal}{arXiv preprint arXiv:2007.07896}  (\bibinfo{year}{2020}).

\bibitem[{\citenamefont{Gaiotto et~al.}(2015)\citenamefont{Gaiotto, Kapustin,
  Seiberg, and Willett}}]{gaiotto2015generalized}
\bibinfo{author}{\bibfnamefont{D.}~\bibnamefont{Gaiotto}},
  \bibinfo{author}{\bibfnamefont{A.}~\bibnamefont{Kapustin}},
  \bibinfo{author}{\bibfnamefont{N.}~\bibnamefont{Seiberg}}, \bibnamefont{and}
  \bibinfo{author}{\bibfnamefont{B.}~\bibnamefont{Willett}},
  \bibinfo{journal}{Journal of High Energy Physics}
  \textbf{\bibinfo{volume}{2015}}, \bibinfo{pages}{172} (\bibinfo{year}{2015}).

\bibitem[{\citenamefont{Else and Senthil}(to appear)}]{else2020qlm}
\bibinfo{author}{\bibfnamefont{D.~V.} \bibnamefont{Else}} \bibnamefont{and}
  \bibinfo{author}{\bibfnamefont{T.}~\bibnamefont{Senthil}} (\bibinfo{year}{to
  appear}).

\bibitem[{\citenamefont{Mross and Senthil}(2011)}]{mross2011decohering}
\bibinfo{author}{\bibfnamefont{D.~F.} \bibnamefont{Mross}} \bibnamefont{and}
  \bibinfo{author}{\bibfnamefont{T.}~\bibnamefont{Senthil}},
  \bibinfo{journal}{Physical Review B} \textbf{\bibinfo{volume}{84}},
  \bibinfo{pages}{165126} (\bibinfo{year}{2011}).

\bibitem[{\citenamefont{Pisarski and Tsvelik}(2021)}]{moat}
\bibinfo{author}{\bibfnamefont{R.~D.} \bibnamefont{Pisarski}} \bibnamefont{and}
  \bibinfo{author}{\bibfnamefont{A.~M.} \bibnamefont{Tsvelik}},
  \bibinfo{journal}{arXiv preprint arXiv:2103.15835}  (\bibinfo{year}{2021}).

\bibitem[{\citenamefont{Polchinski}(1992)}]{polchinski1992effective}
\bibinfo{author}{\bibfnamefont{J.}~\bibnamefont{Polchinski}},
  \bibinfo{journal}{arXiv preprint hep-th/9210046}  (\bibinfo{year}{1992}).

\bibitem[{\citenamefont{Lake}(to appear)}]{lake2021fermi}
\bibinfo{author}{\bibfnamefont{E.}~\bibnamefont{Lake}} (\bibinfo{year}{to
  appear}).

\bibitem[{\citenamefont{Kohn and Luttinger}(1965)}]{kohn1965new}
\bibinfo{author}{\bibfnamefont{W.}~\bibnamefont{Kohn}} \bibnamefont{and}
  \bibinfo{author}{\bibfnamefont{J.}~\bibnamefont{Luttinger}},
  \bibinfo{journal}{Physical Review Letters} \textbf{\bibinfo{volume}{15}},
  \bibinfo{pages}{524} (\bibinfo{year}{1965}).

\bibitem[{\citenamefont{Scalapino et~al.}(1993)\citenamefont{Scalapino, White,
  and Zhang}}]{scalapino1993insulator}
\bibinfo{author}{\bibfnamefont{D.~J.} \bibnamefont{Scalapino}},
  \bibinfo{author}{\bibfnamefont{S.~R.} \bibnamefont{White}}, \bibnamefont{and}
  \bibinfo{author}{\bibfnamefont{S.}~\bibnamefont{Zhang}},
  \bibinfo{journal}{Physical Review B} \textbf{\bibinfo{volume}{47}},
  \bibinfo{pages}{7995} (\bibinfo{year}{1993}).

\bibitem[{\citenamefont{Resta}(2018)}]{resta2018drude}
\bibinfo{author}{\bibfnamefont{R.}~\bibnamefont{Resta}},
  \bibinfo{journal}{Journal of Physics: Condensed Matter}
  \textbf{\bibinfo{volume}{30}}, \bibinfo{pages}{414001}
  (\bibinfo{year}{2018}).

\bibitem[{\citenamefont{Pfleiderer et~al.}(1997)\citenamefont{Pfleiderer,
  McMullan, Julian, and Lonzarich}}]{pfleiderer1997magnetic}
\bibinfo{author}{\bibfnamefont{C.}~\bibnamefont{Pfleiderer}},
  \bibinfo{author}{\bibfnamefont{G.}~\bibnamefont{McMullan}},
  \bibinfo{author}{\bibfnamefont{S.}~\bibnamefont{Julian}}, \bibnamefont{and}
  \bibinfo{author}{\bibfnamefont{G.}~\bibnamefont{Lonzarich}},
  \bibinfo{journal}{Physical Review B} \textbf{\bibinfo{volume}{55}},
  \bibinfo{pages}{8330} (\bibinfo{year}{1997}).

\bibitem[{\citenamefont{Pfleiderer et~al.}(2004)\citenamefont{Pfleiderer,
  Reznik, Pintschovius, L{\"o}hneysen, Garst, and
  Rosch}}]{pfleiderer2004partial}
\bibinfo{author}{\bibfnamefont{C.}~\bibnamefont{Pfleiderer}},
  \bibinfo{author}{\bibfnamefont{D.}~\bibnamefont{Reznik}},
  \bibinfo{author}{\bibfnamefont{L.}~\bibnamefont{Pintschovius}},
  \bibinfo{author}{\bibfnamefont{H.~v.} \bibnamefont{L{\"o}hneysen}},
  \bibinfo{author}{\bibfnamefont{M.}~\bibnamefont{Garst}}, \bibnamefont{and}
  \bibinfo{author}{\bibfnamefont{A.}~\bibnamefont{Rosch}},
  \bibinfo{journal}{Nature} \textbf{\bibinfo{volume}{427}},
  \bibinfo{pages}{227} (\bibinfo{year}{2004}).

\bibitem[{\citenamefont{Muhlbauer et~al.}(2009)\citenamefont{Muhlbauer, Binz,
  Jonietz, Pfleiderer, Rosch, Neubauer, Georgii, and Boni}}]{Muhlbauer_2009}
\bibinfo{author}{\bibfnamefont{S.}~\bibnamefont{Muhlbauer}},
  \bibinfo{author}{\bibfnamefont{B.}~\bibnamefont{Binz}},
  \bibinfo{author}{\bibfnamefont{F.}~\bibnamefont{Jonietz}},
  \bibinfo{author}{\bibfnamefont{C.}~\bibnamefont{Pfleiderer}},
  \bibinfo{author}{\bibfnamefont{A.}~\bibnamefont{Rosch}},
  \bibinfo{author}{\bibfnamefont{A.}~\bibnamefont{Neubauer}},
  \bibinfo{author}{\bibfnamefont{R.}~\bibnamefont{Georgii}}, \bibnamefont{and}
  \bibinfo{author}{\bibfnamefont{P.}~\bibnamefont{Boni}},
  \bibinfo{journal}{Science} \textbf{\bibinfo{volume}{323}},
  \bibinfo{pages}{915–919} (\bibinfo{year}{2009}), ISSN
  \bibinfo{issn}{1095-9203},
  \urlprefix\url{http://dx.doi.org/10.1126/science.1166767}.

\bibitem[{\citenamefont{{Pfleiderer} et~al.}(2001)\citenamefont{{Pfleiderer},
  {Julian}, and {Lonzarich}}}]{NFLMnSi}
\bibinfo{author}{\bibfnamefont{C.}~\bibnamefont{{Pfleiderer}}},
  \bibinfo{author}{\bibfnamefont{S.~R.} \bibnamefont{{Julian}}},
  \bibnamefont{and} \bibinfo{author}{\bibfnamefont{G.~G.}
  \bibnamefont{{Lonzarich}}}, \bibinfo{journal}{\nat}
  \textbf{\bibinfo{volume}{414}}, \bibinfo{pages}{427} (\bibinfo{year}{2001}).

\bibitem[{\citenamefont{Nagaosa and Tokura}(2013)}]{nagaosaMnSi}
\bibinfo{author}{\bibfnamefont{N.}~\bibnamefont{Nagaosa}} \bibnamefont{and}
  \bibinfo{author}{\bibfnamefont{Y.}~\bibnamefont{Tokura}},
  \bibinfo{journal}{Nature nanotechnology} \textbf{\bibinfo{volume}{8}},
  \bibinfo{pages}{899} (\bibinfo{year}{2013}).

\bibitem[{\citenamefont{Pedrazzini et~al.}(2007)\citenamefont{Pedrazzini,
  Wilhelm, Jaccard, Jarlborg, Schmidt, Hanfland, Akselrud, Yuan, Schwarz, Grin
  et~al.}}]{FeGeNFL}
\bibinfo{author}{\bibfnamefont{P.}~\bibnamefont{Pedrazzini}},
  \bibinfo{author}{\bibfnamefont{H.}~\bibnamefont{Wilhelm}},
  \bibinfo{author}{\bibfnamefont{D.}~\bibnamefont{Jaccard}},
  \bibinfo{author}{\bibfnamefont{T.}~\bibnamefont{Jarlborg}},
  \bibinfo{author}{\bibfnamefont{M.}~\bibnamefont{Schmidt}},
  \bibinfo{author}{\bibfnamefont{M.}~\bibnamefont{Hanfland}},
  \bibinfo{author}{\bibfnamefont{L.}~\bibnamefont{Akselrud}},
  \bibinfo{author}{\bibfnamefont{H.~Q.} \bibnamefont{Yuan}},
  \bibinfo{author}{\bibfnamefont{U.}~\bibnamefont{Schwarz}},
  \bibinfo{author}{\bibfnamefont{Y.}~\bibnamefont{Grin}}, \bibnamefont{et~al.},
  \bibinfo{journal}{Phys. Rev. Lett.} \textbf{\bibinfo{volume}{98}},
  \bibinfo{pages}{047204} (\bibinfo{year}{2007}),
  \urlprefix\url{https://link.aps.org/doi/10.1103/PhysRevLett.98.047204}.

\bibitem[{\citenamefont{Bak and Jensen}(1980)}]{bak1980theory}
\bibinfo{author}{\bibfnamefont{P.}~\bibnamefont{Bak}} \bibnamefont{and}
  \bibinfo{author}{\bibfnamefont{M.~H.} \bibnamefont{Jensen}},
  \bibinfo{journal}{Journal of Physics C: Solid State Physics}
  \textbf{\bibinfo{volume}{13}}, \bibinfo{pages}{L881} (\bibinfo{year}{1980}).

\bibitem[{\citenamefont{Nakanishi et~al.}(1980)\citenamefont{Nakanishi, Yanase,
  Hasegawa, and Kataoka}}]{nakanishi1980origin}
\bibinfo{author}{\bibfnamefont{O.}~\bibnamefont{Nakanishi}},
  \bibinfo{author}{\bibfnamefont{A.}~\bibnamefont{Yanase}},
  \bibinfo{author}{\bibfnamefont{A.}~\bibnamefont{Hasegawa}}, \bibnamefont{and}
  \bibinfo{author}{\bibfnamefont{M.}~\bibnamefont{Kataoka}},
  \bibinfo{journal}{Solid State Communications} \textbf{\bibinfo{volume}{35}},
  \bibinfo{pages}{995} (\bibinfo{year}{1980}).

\bibitem[{\citenamefont{Hopkinson and Kee}(2009)}]{hopkinson2009origin}
\bibinfo{author}{\bibfnamefont{J.~M.} \bibnamefont{Hopkinson}}
  \bibnamefont{and} \bibinfo{author}{\bibfnamefont{H.-Y.} \bibnamefont{Kee}},
  \bibinfo{journal}{Physical Review B} \textbf{\bibinfo{volume}{79}},
  \bibinfo{pages}{014421} (\bibinfo{year}{2009}).

\bibitem[{\citenamefont{Seiberg and
  Shao}(2020{\natexlab{c}})}]{seiberg2020exoticIII}
\bibinfo{author}{\bibfnamefont{N.}~\bibnamefont{Seiberg}} \bibnamefont{and}
  \bibinfo{author}{\bibfnamefont{S.-H.} \bibnamefont{Shao}},
  \bibinfo{journal}{arXiv preprint arXiv:2004.06115}
  (\bibinfo{year}{2020}{\natexlab{c}}).

\bibitem[{\citenamefont{Gorantla et~al.}(2020)\citenamefont{Gorantla, Lam,
  Seiberg, and Shao}}]{seiberg2020moreexotic}
\bibinfo{author}{\bibfnamefont{P.}~\bibnamefont{Gorantla}},
  \bibinfo{author}{\bibfnamefont{H.~T.} \bibnamefont{Lam}},
  \bibinfo{author}{\bibfnamefont{N.}~\bibnamefont{Seiberg}}, \bibnamefont{and}
  \bibinfo{author}{\bibfnamefont{S.-H.} \bibnamefont{Shao}},
  \bibinfo{journal}{arXiv preprint arXiv:2007.04904}  (\bibinfo{year}{2020}).

\end{thebibliography}

\end{document}